\documentclass[fleqn,usenatbib]{mnras}

\usepackage{newtxtext,newtxmath}

\usepackage[T1]{fontenc}

\DeclareRobustCommand{\VAN}[3]{#2}
\let\VANthebibliography\thebibliography
\def\thebibliography{\DeclareRobustCommand{\VAN}[3]{##3}\VANthebibliography}

\usepackage{graphicx}	
\usepackage{amsmath}	
\usepackage{xcolor}

\title[WD mass growth in CVs: Roles of DNe]{White Dwarf Mass Growth in Cataclysmic Variables: Roles of Dwarf Novae}

\author[Wei-Min Liu et al.]{
Wei-Min Liu,$^{1}$\thanks{E-mail: liuwmph@163.com}
Long Jiang,$^{2}$
Wen-Cong Chen$^{2}$\thanks{E-mail: chenwc@pku.edu.cn}
and Xiang-Dong Li$^{3,4}$\thanks{E-mail: lixd@nju.edu.cn}
\\
$^{1}$School of Physics and Electrical Information, Shangqiu Normal University, Shangqiu 476000, China\\
$^{2}$School of Science, Qingdao University of Technology, Qingdao 266525, China\\
$^{3}$School of Astronomy and Space Science, Nanjing University, Nanjing 210023, China\\
$^4$Key Laboratory of Modern Astronomy and Astrophysics (Nanjing University), Ministry of Education, Nanjing 210023, China}

\date{Accepted XXX. Received YYY; in original form ZZZ}

\pubyear{2023}

\begin{document}
\label{firstpage}
\pagerange{\pageref{firstpage}--\pageref{lastpage}}
\maketitle

\begin{abstract}
The disc instability mechanism (DIM) is widely accepted to account for the transient behaviour of dwarf novae (DNe), which experience short outbursts separated by long quiescence. The duty cycle (the ratio between the outburst duration and the recurrence time) determines the amount of accreted mass by the white dwarf (WDs) during outbursts, thus playing an important role in the long-term binary evolution. Employing the code of Modules for Experiments in Stellar Astrophysics, we systemically investigate the influence of the duty cycles on the evolution of DNe and the mass growth of accreting carbon-oxygen (CO) WDs. Our calculations show that, while the DIM can considerably influence the accretion process, efficient WD-mass growth requires a particular range of the duty cycle. For WDs with the initial masses of 0.6, 0.7 and 1.1 $M_\odot$, these duty cycles are 0.006$\,\leq$$d$$\,\leq$0.007, $d$\,=\,0.005 and $d$\,=\,0.003, and the accumulated mass of the WDs can reach 0.1, 0.13 and 0.21 $M_\odot$, respectively. In all of our simulations, no CO WDs can grow their masses to the explosion mass of Type Ia supernovae of about $1.38~M_\odot$. Because of a much short timescale of the outburst state, the final donor-star masses and orbital periods are insensitive to the duty cycles. Therefore, we propose that the DIM in DNe could alleviate the WD mass problem to some extent.
\end{abstract}




\section{Introduction}
Cataclysmic variables (CVs) are short-period interacting binaries in which a low-mass donor star is transferring materials to a carbon-oxygen (CO) white dwarf \citep[WDs, see][for reviews]{war95,rit10,kni11}. Angular momentum loss (AML) plays a crucial role in the secular evolution of CVs. It is traditionally thought that, above the period gap (orbital periods between 2 and 3 hours), magnetic braking \citep[MB,][]{verb81,Rappaport1983} dominates the evolution of CVs, while gravitational radiation \citep[GR,][]{landau75} is solely responsible for AML below the period gap   because MB ceases for a fully convective donor star \citep{para17}. However, it still remains controversial whether MB really completely disappears in this situation \citep{pat98,kni11,sark22}.

Dwarf novae (DNe) are a subtype of CVs which switch between long-term quiescence and short-term outbursts \citep{osa74}. According to the thermal-viscous disc instability mechanism \citep[DIM,][]{osa96,hel01}, the mass-transfer rate from the donor star during quiescence is much greater than that from the disc onto the WD. Therefore, the transferred material piles up in the disc and the temperature of the disc rises owing to viscous heating. When the disc material becomes fully ionised, a sudden mass transfer onto the WD occurs, leading to an outburst. 

Considering the influence of DIM on the mass-transfer process, \cite{king03} presented an alternative evolutionary channel for long-period DNe, and showed that a WD with initial mass $~0.7 M_\odot$ could grow to the explosion mass of Type Ia supernovae (SNe Ia, about 1.38 $M_\odot$) if the system experiences thermal-timescale mass transfer in the early evolutionary stages. Adopting the similar idea, \cite{xu09} investigated accretion-induced collapse of WDs in long-period DNe, and suggested that the evolution can well account for the formation of the bursting pulsar GRO J1744-28 if the duty cycle (i.e., the ratio between the outburst duration and the recurrence time) is $d=0.003$. \cite{wang10} explored in detail the formation of long delay-time SNe Ia taking into account DIM (with $d=0.01$). They showed that it is possible for a 0.6 $M_\odot$ WD to grow to the Chandrasekhar limit with an initial donor star of mass less than 1.7 $M_\odot$. Subsequent studies on the evolution of WD binaries \citep{chen14,toon14,liud19}, neutron star X-ray binaries \citep{liu11,jia16,gao22}, and black hole X-ray binaries \citep{shao20} also adopted constant value of $d$. However, the derived $d$ from observations of DNe range from a few $10^{-3}$ to around 0.5, which might be related to the orbital properties of the binaries \citep{bri15}. Since the accretion behaviour of the WD during outbursts sensitively depend on the magnitude of $d$, using single value of $d$ is obviously a too simplified assumption.

Our purpose is to systematically model the evolutionary processes of DNe that experience the accretion-disc instability, and investigate the influence of the initial parameters including the initial WD mass, initial donor-star mass, initial orbital period, and duty cycle on the evolution of DNe. Meanwhile, we hope to contribute an evolutionary channel to solve the WD mass problem. In observations, the average mass of isolated WDs is about 0.6 $M_\odot$ \citep{kep07}, while the mean mass of WDs in CVs are obviously more massive (about $0.8~M_\odot$) \citep{zor11}. \cite{sch16} argued that this mass problem can be solved by consequential AML, but the subsequent study indicated that the proposed physical interpretation of this empirical consequential AML still need to be improved \citep{liu19}. As mentioned above, both \cite{king03} and \cite{xu09} present an evolutionary example with a low duty cycle ($d\approx0.003$). An observational study of dwarf nova AX J1549.8--5416 implied that a duty cycle of 0.5 \citep{zhang17}. Therefore, the duty cycles might have a wide distribution,  $d=0.01$ used in the investigation on the progenitors of SNe Ia is probably implausible for the evolution of DNe. A thorough research for the influence of duty cycles on the evolution of DNe is required.

In Section 2, we describe the input physics and binary evolution code. The numerically calculated results are presented in Section 3. Sections 4, and 5 give a brief discussion and conclusion, respectively.

\section{Input physics and binary evolution code}
In this section, we firstly introduce the parameterised condition for DIMs of DNe, and a critical mass-transfer rate under which DIMs occur. Subsequently, the mass-growth law of WDs during the hydrogen and helium burning, and detailed settings in the binary evolutionary code are presented.

\subsection{Mass Transfer and Accretion Disc Instability}
By parameterising the properties of DNe, \cite{smak83} found that the accretion in WD binaries is unstable if the mass-transfer rate ($\dot{M}_{\rm d}$) from the donor star is less than a critical mass-transfer rate $\dot{M}_{\rm {d,cr}}$,  which is related with the orbital period. Actually, this critical mass-transfer rate depends on the effective temperature at the outer edge of the accretion disc. Because of the thermal and viscous instability of the accretion disc, the accretion flow is unstable if the effective temperature is below a critical temperature as
\begin{equation}
T_{\rm cr}\approx 7943(R_{\rm disc}/10^{10}~ {\rm cm})^{-0.1}~ {\rm K},
\end{equation}
where $R_{\rm disc}$ is the outer edge radius of the accretion disc \citep{smak83}. This critical temperature is consistent with the hydrogen ionisation temperature of about 6500 K \citep{King1997}.  Corresponding to the critical temperature, the critical mass-transfer rate can be written as \citep{VanPara1996}
\begin{equation}
\dot{M}_{\rm {d,cr}}=4.3\times10^{-9}(P_{\rm orb}/4\,{\rm hr})^{1.7}{\,M_\odot\,\rm yr}^{-1},
\end{equation}
where $P_{\rm orb}$ is the orbital period. 

If the mass-transfer rate from the donor star is less than $\dot{M}_{\rm {d,cr}}$, the accretion disc would suffer thermal and viscous instability. Once the accretion disc is unstable, the mass accretion on to the WD would transit between the short outburst state and the long quiescent state. Given a recurrence time $t_{\rm rec}$, the duty cycle is defined as $d=t_{\rm out}/t_{\rm rec}$, where $t_{\rm out}$ is the duration of the outburst state. The mass accumulated in the accretion disc within the quiescent time (which is close to the recurrence time when $d\ll1$) is assumed to be accreted by the WD in the outburst stage, i.e. ${|\dot M}_{\rm d}|\times t_{\rm rec}={\dot M}_{\rm acc}\times t_{\rm out}$. Due to the disc instability, the accreting WD has a chance to experience a steady mass accumulation,
resulting in an efficient mass growth of the accreting WD \citep{kah97}. Therefore, we take an accretion rate of the WD as follows
\begin{equation}
\dot{M}_{\rm acc}= \left\{\begin{array}{l@{\quad}l} |\dot{M}_{\rm d}|/d,
& |\dot{M}_{\rm d}|\leq\dot{M}_{\rm d,cr} \strut\\
|\dot{M}_{\rm d}|
, & |\dot{M}_{\rm d}|>\dot{M}_{\rm d,cr} \strut\\\end{array}\right.
\end{equation}
It is noteworthy that the WD only accretes mass from the accretion disc at a rate of $\dot{M}_{\rm acc}$ during the outburst state once the disc instability occurs.

\subsection{Mass Growth of WDs}
During the accretion of the WD, its mass growth depends on the mass accumulation efficiencies during the hydrogen and helium burning stages. Similar to \cite{liu19}, the mass-growth rate of a WD is given by
\begin{equation}
\dot{M}_{\rm WD}=\eta_{\rm H}\eta_{\rm He}{\dot M}_{\rm acc},
\end{equation}
where $\eta_{\rm H}$ and $\eta_{\rm He}$ are the mass accumulation efficiencies for hydrogen and helium burning, respectively. We follow \cite{hil16} and \cite{Ka2004} for the description of $\eta_{\rm H}$ and $\eta_{\rm He}$, which are determined by the WD mass and the accretion rate \citep[see also][for more details]{liu16}. There exists an accretion-rate range for the steady burning of the accreting hydrogen. The upper critical accretion rate for steady hydrogen burning can be written as \citep{Han04}
\begin{equation}
\dot{M}_{\rm {H,up}}=5.3\times10^{-7}\left(\frac{1.7-X}{X}\right)(M_{\rm WD}/{M_\odot}-0.4){\,M_\odot\,\rm yr}^{-1},
\end{equation}
where $X$ is the hydrogen abundance of the donor star. When $\dot{M}_{\rm acc}$ is greater than $\dot{M}_{\rm {H,up}}$, the accreted Hydrogen
stably burns into He at a rate of $\dot{M}_{\rm {H,up}}$, and the excess material is blown away in the form of the optically thick winds \citep{ka1994}. Furthermore, if the mass-accretion rate for the hydrogen material is lower than $\dot{M}_{\rm H,low}$\,=\,$3\times 10^{-8}\,M_\odot\,{\rm yr}^{-1}$, strong hydrogen-shell flashes take place, and all accreted material is ejected from the surface of the WD \citep{kove94}. During the hydrogen and helium burning, the excess material in unit time (${\dot M}_{\rm acc}-{\dot M}_{\rm WD}$) is thought to  carry away the specific AM of the WD. 

\subsection{Binary Evolutionary Code}
We carried out binary evolutionary calculations of DNe using the Modules for Experiments in Stellar Astrophysics \citep[MESA,][]{Paxton2015ApJS}. At the beginning of the model, a binary system containing a WD and an MS companion star is assumed to exist in a circular and synchronised orbit. In the calculations, the WD is thought to be a point mass. For the chemical abundance of the donor star, we take a solar composition ($X = 0.70$, $Y = 0.28$ and $Z = 0.02$). Our simulations depend on four initial input parameters including the initial donor-star mass $M_{\rm {d,i}}$, initial WD mass $M_{\rm {WD,i}}$, initial orbital period $P_{\rm {orb,i}}$, and duty cycle $d$. 

Except for AML by the mass loss mentioned in Section 2.2, we consider systematic AML including the MB above the period gap and GR \citep{landau75}. For MB mechanism, we adopt the standard MB description given by \cite{Rappaport1983}, and take the MB index as $\gamma=4$ \citep{verb81}.

\section{Calculated Results}

\begin{table}
\begin{center}
\caption{Initial parameters and the final WD masses for $M_{\rm WD,i}=0.6~M_\odot$}
\centering
\begin{tabular}{ccccccc}
\hline\hline
$M_{\rm d,i}/M_\odot$ & $M_{\rm {WD,i}}/M_\odot$ &$M_{\rm {WD,f}}/M_\odot$ &$P_{\rm {orb,i}}/\rm d$ & duty cycle \\
\hline\hline
0.5   &0.6   &0.634   &0.5   &0.002   \\
0.5   &0.6   &0.672   &0.5   &0.005   \\
0.5   &0.6   &0.660   &0.5   &0.01   \\
0.5   &0.6   &0.608   &0.5   &0.05   \\
0.5   &0.6   &0.602   &0.5   &0.1   \\
\hline
0.5   &0.6   &0.635   &0.79   &0.002   \\
0.5   &0.6   &0.676   &0.79   &0.005   \\
0.5   &0.6   &0.663   &0.79   &0.01   \\
0.5   &0.6   &0.609   &0.79   &0.05   \\
0.5   &0.6   &0.602   &0.79   &0.1   \\
\hline
0.5   &0.6   &0.636   &1.0   &0.002   \\
0.5   &0.6   &0.678   &1.0   &0.005   \\
0.5   &0.6   &0.665   &1.0   &0.01   \\
0.5   &0.6   &0.609   &1.0   &0.05   \\
0.5   &0.6   &0.602   &1.0   &0.1   \\
\hline\hline
0.6   &0.6   &0.640   &0.5   &0.002   \\
0.6   &0.6   &0.686   &0.5   &0.005   \\
0.6   &0.6   &0.680   &0.5   &0.01   \\
0.6   &0.6   &0.614   &0.5   &0.05   \\
0.6   &0.6   &0.604   &0.5   &0.1   \\
\hline
0.6   &0.6   &0.643   &1.0   &0.002   \\
0.6   &0.6   &0.692   &1.0   &0.005   \\
0.6   &0.6   &0.687   &1.0   &0.01   \\
0.6   &0.6   &0.616   &1.0   &0.05   \\
0.6   &0.6   &0.605   &1.0   &0.1   \\
\hline
0.6   &0.6   &0.646   &1.26   &0.002   \\
0.6   &0.6   &0.696   &1.26   &0.005   \\
0.6   &0.6   &0.691   &1.26   &0.01   \\
0.6   &0.6   &0.618   &1.26   &0.05   \\
0.6   &0.6   &0.605   &1.26   &0.1   \\
\hline\hline
0.7   &0.6   &0.660   &0.5   &0.002   \\
0.7   &0.6   &0.686   &0.5   &0.005   \\
0.7   &0.6   &0.687   &0.5   &0.01   \\
0.7   &0.6   &0.614   &0.5   &0.05   \\
0.7   &0.6   &0.604   &0.5   &0.1   \\
\hline
0.7   &0.6   &0.644   &1.0   &0.002   \\
0.7   &0.6   &0.694   &1.0   &0.005   \\
0.7   &0.6   &0.692   &1.0   &0.01   \\
0.7   &0.6   &0.618   &1.0   &0.05   \\
0.7   &0.6   &0.605   &1.0   &0.1   \\
\hline
0.7   &0.6   &0.651   &1.58   &0.002   \\
0.7   &0.6   &0.694   &1.58   &0.005   \\
0.7   &0.6   &0.700   &1.58   &0.01   \\
0.7   &0.6   &0.618   &1.58   &0.05   \\
0.7   &0.6   &0.605   &1.58   &0.1   \\
\hline\hline
\end{tabular}
\end{center}
\end{table}

\begin{table}
\begin{center}
\caption{Same as Table 1 but for $M_{\rm WD,i}=0.7~M_\odot$}
\centering
\begin{tabular}{ccccccc}
\hline\hline
$M_{\rm d,i}/M_\odot$ & $M_{\rm {WD,i}}/M_\odot$ &$M_{\rm {WD,f}}/M_\odot$ &$P_{\rm {orb,i}}/\rm d$ &  duty cycle \\
\hline\hline
0.7   &0.7   &0.784   &0.5   &0.002   \\
0.7   &0.7   &0.818   &0.5   &0.005   \\
0.7   &0.7   &0.800   &0.5   &0.01   \\
0.7   &0.7   &0.726   &0.5   &0.05   \\
0.7   &0.7   &0.708   &0.5   &0.1   \\
\hline
0.7   &0.7   &0.791   &1.0   &0.002   \\
0.7   &0.7   &0.824   &1.0   &0.005   \\
0.7   &0.7   &0.800   &1.0   &0.01   \\
0.7   &0.7   &0.730   &1.0   &0.05   \\
0.7   &0.7   &0.709   &1.0   &0.1   \\
\hline
0.7   &0.7   &0.802   &1.58   &0.002   \\
0.7   &0.7   &0.831   &1.58   &0.005   \\
0.7   &0.7   &0.800   &1.58   &0.01   \\
0.7   &0.7   &0.734   &1.58   &0.05   \\
0.7   &0.7   &0.712   &1.58   &0.1   \\
\hline\hline
0.8   &0.7   &0.784  &0.5   &0.002   \\
0.8   &0.7   &0.819   &0.5   &0.005   \\
0.8   &0.7   &0.800   &0.5   &0.01   \\
0.8   &0.7   &0.726   &0.5   &0.05   \\
0.8   &0.7   &0.706   &0.5   &0.1   \\
\hline
0.8   &0.7   &0.784   &1.0   &0.002   \\
0.8   &0.7   &0.822   &1.0   &0.005   \\
0.8   &0.7   &0.800   &1.0   &0.01   \\
0.8   &0.7   &0.728   &1.0   &0.05   \\
0.8   &0.7   &0.708   &1.0   &0.1   \\
\hline
0.8   &0.7   &0.804   &1.58   &0.002   \\
0.8   &0.7   &0.826   &1.58   &0.005   \\
0.8   &0.7   &0.800   &1.58   &0.01   \\
0.8   &0.7   &0.730   &1.58   &0.05   \\
0.8   &0.7   &0.709   &1.58   &0.1   \\
\hline\hline
0.9   &0.7   &0.784   &0.5   &0.002   \\
0.9   &0.7   &0.819   &0.5   &0.005   \\
0.9   &0.7   &0.800   &0.5   &0.01   \\
0.9   &0.7   &0.727   &0.5   &0.05   \\
0.9   &0.7   &0.708   &0.5   &0.1   \\
\hline
0.9   &0.7   &0.789   &1.0   &0.002   \\
0.9   &0.7   &0.821   &1.0   &0.005   \\
0.9   &0.7   &0.800   &1.0   &0.01   \\
0.9   &0.7   &0.728   &1.0   &0.05   \\
0.9   &0.7   &0.708   &1.0   &0.1   \\
\hline
0.9   &0.7   &0.806   &1.58   &0.002   \\
0.9   &0.7   &0.826   &1.58   &0.005   \\
0.9   &0.7   &0.802   &1.58   &0.01   \\
0.9   &0.7   &0.732   &1.58   &0.05   \\
0.9   &0.7   &0.710   &1.58   &0.1   \\
\hline\hline
\end{tabular}
\end{center}
\end{table}

\begin{table}
\begin{center}
\caption{Same as Table 1 but for $M_{\rm WD,i}=1.1~M_\odot$}
\centering
\begin{tabular}{ccccccc}
\hline\hline
$M_{\rm d,i}/M_\odot$ & $M_{\rm {WD,i}}/M_\odot$ &$M_{\rm {WD,f}}/M_\odot$ &$P_{\rm {orb,i}}/\rm d$ &  duty cycle \\
\hline\hline
0.8   &1.1   &1.272   &0.5   &0.002   \\
0.8   &1.1   &1.240   &0.5   &0.005   \\
0.8   &1.1   &1.184   &0.5   &0.01   \\
0.8   &1.1   &1.1   &0.5   &0.05   \\
0.8   &1.1   &1.1   &0.5   &0.1   \\
\hline
0.8   &1.1   &1.274   &1.0   &0.002   \\
0.8   &1.1   &1.241   &1.0   &0.005   \\
0.8   &1.1   &1.180   &1.0   &0.01   \\
0.8   &1.1   &1.1   &1.0   &0.05   \\
0.8   &1.1   &1.1   &1.0   &0.1   \\
\hline
0.8   &1.1   &1.278   &1.58   &0.002   \\
0.8   &1.1   &1.240   &1.58   &0.005   \\
0.8   &1.1   &1.159   &1.58   &0.01   \\
0.8   &1.1   &1.1   &1.58   &0.05   \\
0.8   &1.1   &1.1   &1.58   &0.1   \\
\hline\hline
0.9   &1.1   &1.285   &0.5   &0.002   \\
0.9   &1.1   &1.271   &0.5   &0.005   \\
0.9   &1.1   &1.231   &0.5   &0.01   \\
0.9   &1.1   &1.1   &0.5   &0.05   \\
0.9   &1.1   &1.1   &0.5   &0.1   \\
\hline
0.9   &1.1   &1.287   &1.0   &0.002   \\
0.9   &1.1   &1.272   &1.0   &0.005   \\
0.9   &1.1   &1.231   &1.0   &0.01   \\
0.9   &1.1   &1.1   &1.0   &0.05   \\
0.9   &1.1   &1.1   &1.0   &0.1   \\
\hline
0.9   &1.1   &1.292   &1.58   &0.002   \\
0.9   &1.1   &1.273   &1.58   &0.005   \\
0.9   &1.1   &1.230   &1.58   &0.01   \\
0.9   &1.1   &1.1   &1.58   &0.05   \\
0.9   &1.1   &1.1   &1.58   &0.1   \\
\hline\hline
1.0   &1.1   &1.297   &0.5   &0.002   \\
1.0   &1.1   &1.297   &0.5   &0.005   \\
1.0   &1.1   &1.268   &0.5   &0.01   \\
1.0   &1.1   &1.1   &0.5   &0.05   \\
1.0   &1.1   &1.1   &0.5   &0.1   \\
\hline
1.0   &1.1   &1.299   &1.0   &0.002   \\
1.0   &1.1   &1.299   &1.0   &0.005   \\
1.0   &1.1   &1.277   &1.0   &0.01   \\
1.0   &1.1   &1.1   &1.0   &0.05   \\
1.0   &1.1   &1.1   &1.0   &0.1   \\
\hline
1.0   &1.1   &1.309   &1.58   &0.002   \\
1.0   &1.1   &1.316   &1.58   &0.005   \\
1.0   &1.1   &1.278   &1.58   &0.01   \\
1.0   &1.1   &1.1   &1.58   &0.05   \\
1.0   &1.1   &1.1   &1.58   &0.1   \\
\hline\hline
\end{tabular}
\end{center}
\end{table}

\begin{table}
\begin{center}
\caption{The maximum $M_{\rm WD,f}$ and the corresponding duty cycles}
\centering
\begin{tabular}{ccccccc}
\hline\hline
$M_{\rm d,i}/M_\odot$ & $M_{\rm {WD,i}}/M_\odot$ &$M_{\rm {WD,f}}/M_\odot$ &$P_{\rm {orb,i}}/\rm d$ & duty cycle \\
\hline\hline
0.6   &0.6   &0.687   &0.5   &0.006   \\
0.6   &0.6   &0.693   &1.0   &0.006   \\
0.7   &0.6   &0.692   &0.5   &0.007   \\
0.7   &0.6   &0.696   &1.0   &0.007   \\
\hline
0.7   &0.7   &0.818   &0.5   &0.005   \\
0.7   &0.7   &0.831   &1.0   &0.005   \\
0.8   &0.7   &0.819   &0.5   &0.005   \\
0.8   &0.7   &0.822   &1.0   &0.005   \\
\hline
0.9   &1.1   &1.290   &0.5   &0.003   \\
0.9   &1.1   &1.292   &1.0   &0.003   \\
1.0   &1.1   &1.312   &0.5   &0.003   \\
1.0   &1.1   &1.311   &1.0   &0.003   \\
\hline\hline
\end{tabular}
\end{center}
\end{table}

We take the initial WD masses to be $M_{\rm {WD,i}}$\,=\,0.6, 0.7 and 1.1 $M_\odot$, and the corresponding initial donor star masses to be $M_{\rm {d,i}}$\,=\,(0.5, 0.6 and 0.7 $M_\odot$), (0.7, 0.8 and 0.9 $M_\odot$), and (0.8, 0.9 and 1.0 $M_\odot$), respectively. The initial orbital periods are taken to be three typical values in the range from 0.5 to 1.58 d. We model the evolution of 27 WD binaries to investigate the influence of duty cycles on the mass growth of CO WDs in DNe. \cite{king03} derived a duty cycle range from a few times $10^{-3}$ to 0.1 based on the observations \citep{war95}. Therefore, we set $d = 0.002, 0.005, 0.01, 0.05$, and 0.1, which cover the plausible range of duty cycles.

\begin{figure*}
\centering
\includegraphics[scale=0.3,trim={-2cm 2cm 4cm 0cm}]{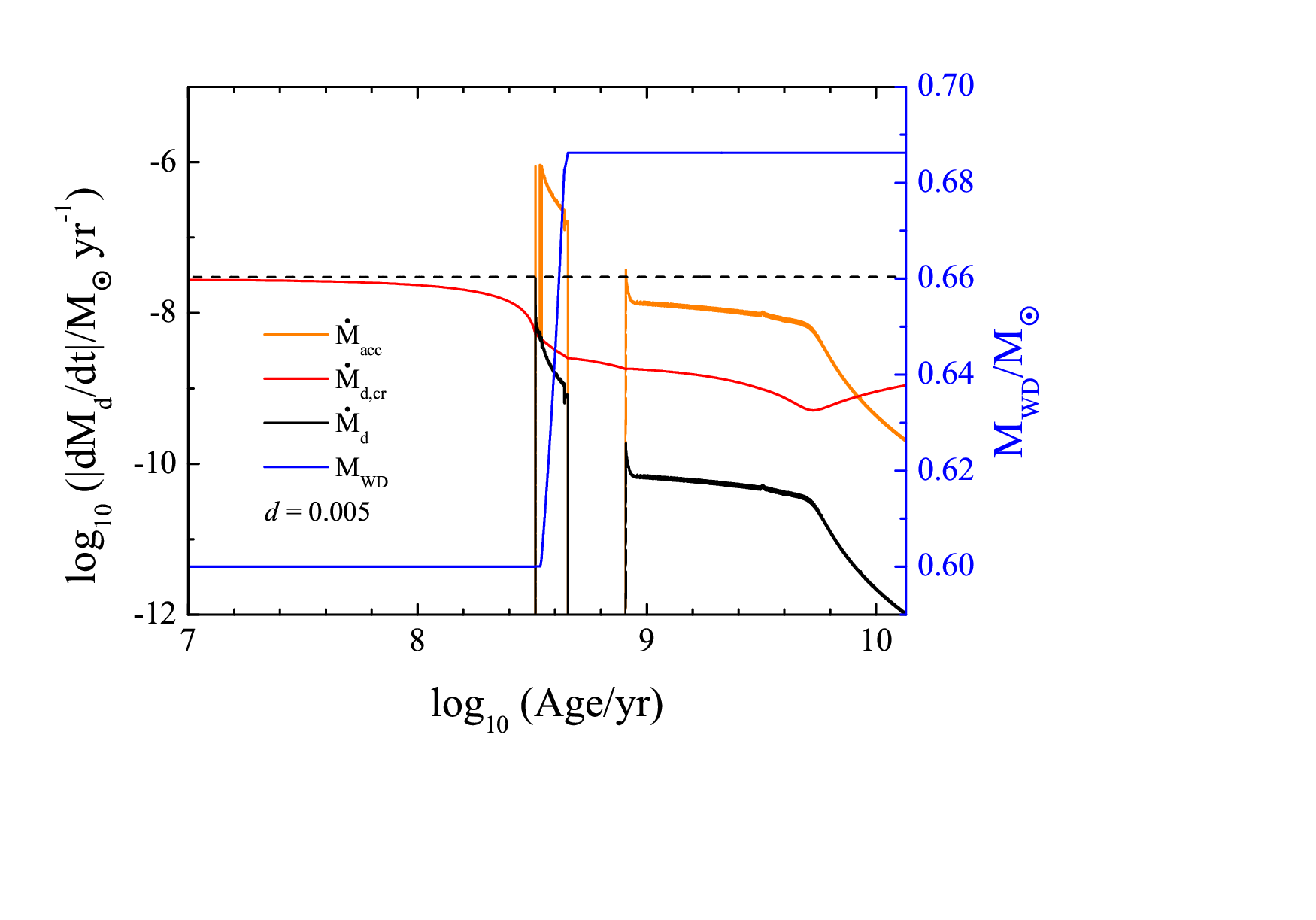}\includegraphics[scale=0.3,trim={2cm 2cm 2cm 0cm}]{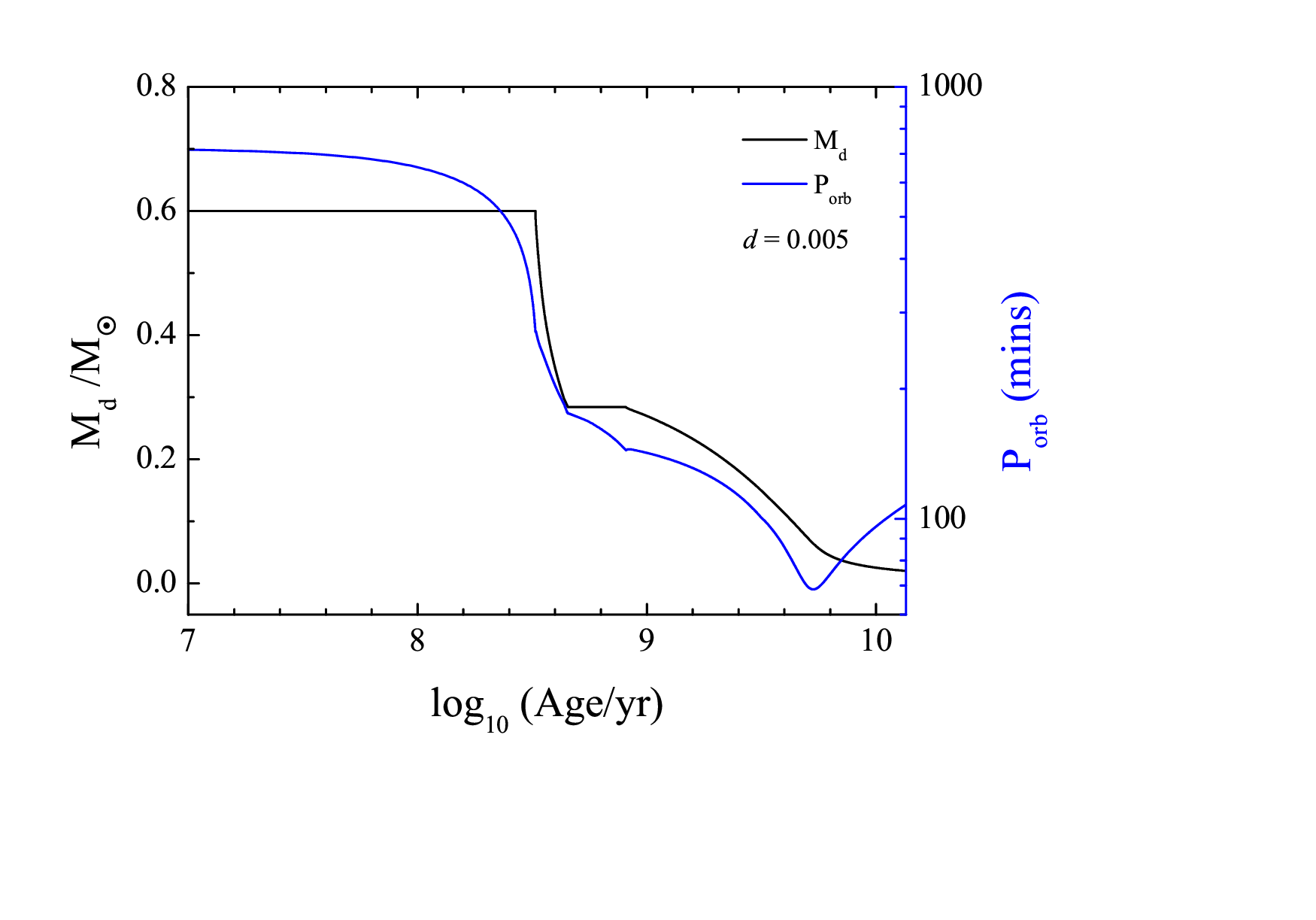}
\includegraphics[scale=0.3,trim={-2cm 0cm 2cm 3cm}]{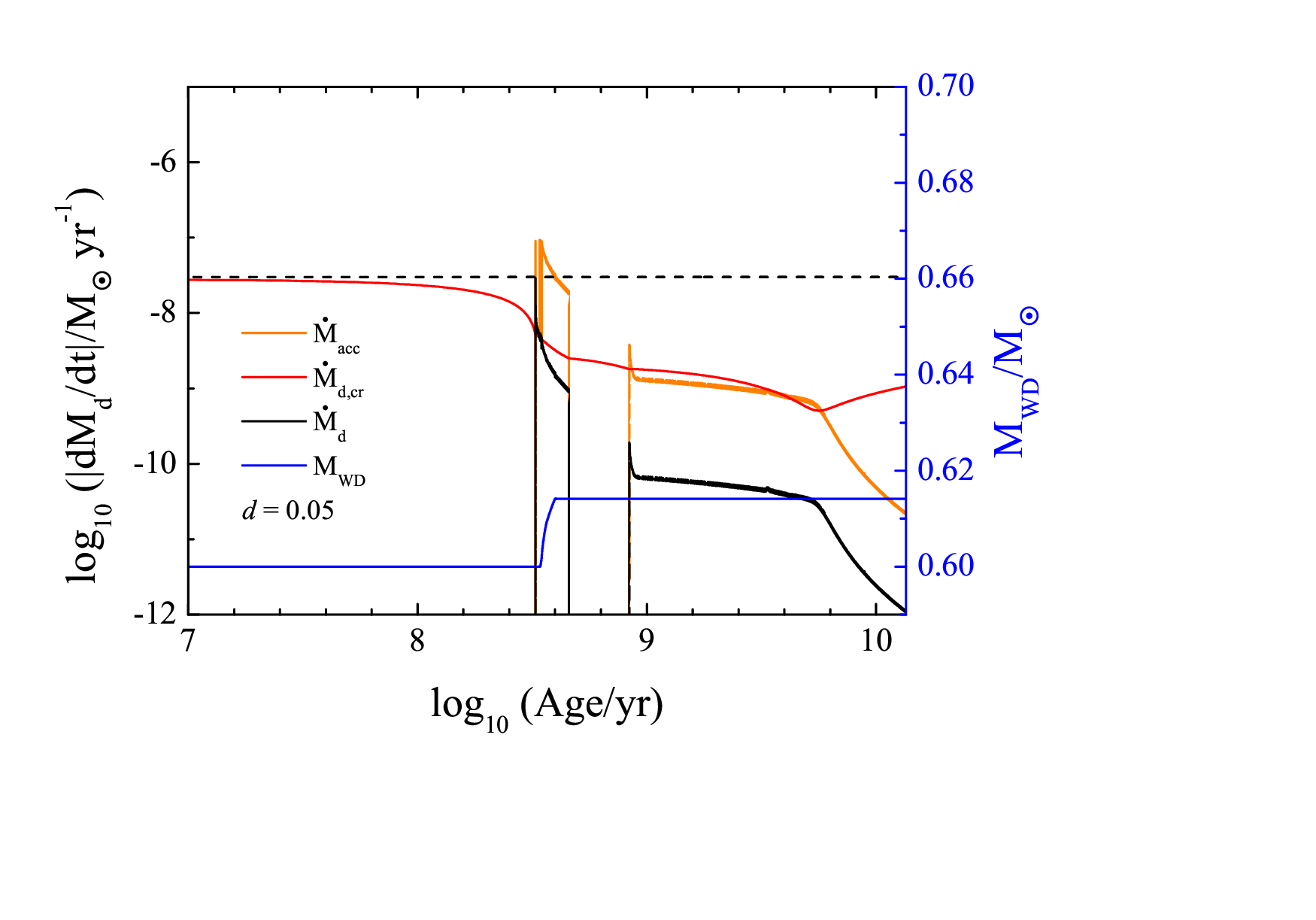}\includegraphics[scale=0.3,trim={4cm 0cm 2cm 3cm}]{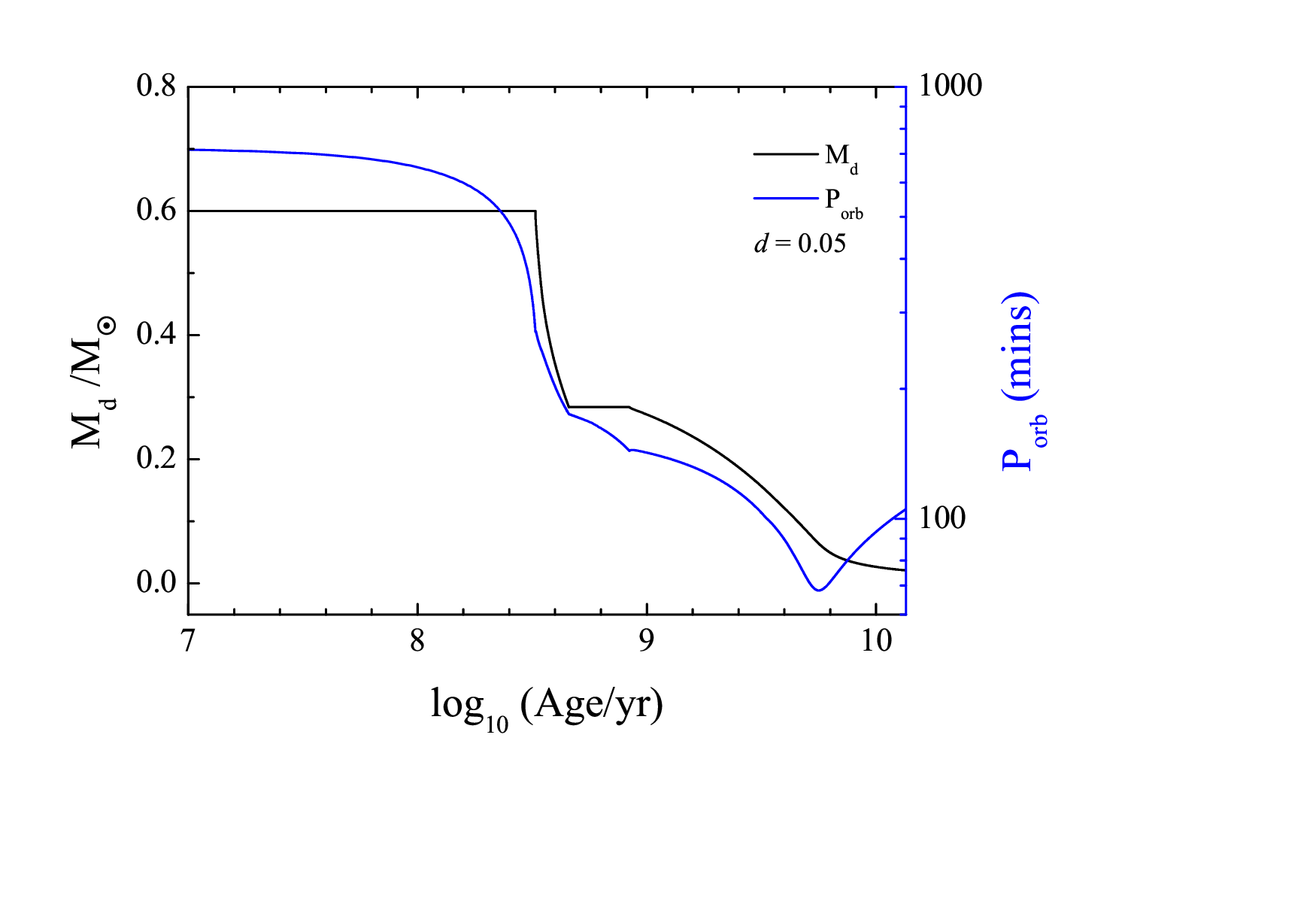}
\vspace{-1cm}
\caption{Evolution of a dwarf nova with $M_{\rm {d,i}}$ = 0.6 $M_\odot$, $M_{\rm {WD,i}}$ = 0.6 $M_\odot$ and $P_{\rm orb,i}$ = 0.5 d for $d=0.005$ (upper panels), and 0.05 (bottom panels).  In the left panels, the orange, red, black, and blue curves represent the evolutionary tracks of the accretion rate $\dot{M}_{\rm acc}$ onto the WD during outburst states, the critical mass-transfer rate $\dot{M}_{\rm d,cr}$ below which DIMs occur, the mass-transfer rate $|\dot{M}_{\rm d}|$ of the donor star, and the WD mass, respectively. The horizontal dashed lines indicate $\dot{M}_{\rm H,low}$ below which strong H-shell flash occurs. In the right panels, the black, and blue curves denote the evolutionary tracks of the donor-star mass, and the orbital period, respectively.}
\label{fig:subfig}
\end{figure*}

\begin{figure*}
\centering
\includegraphics[scale=0.3,trim={-2cm 2cm 4cm 0cm}]{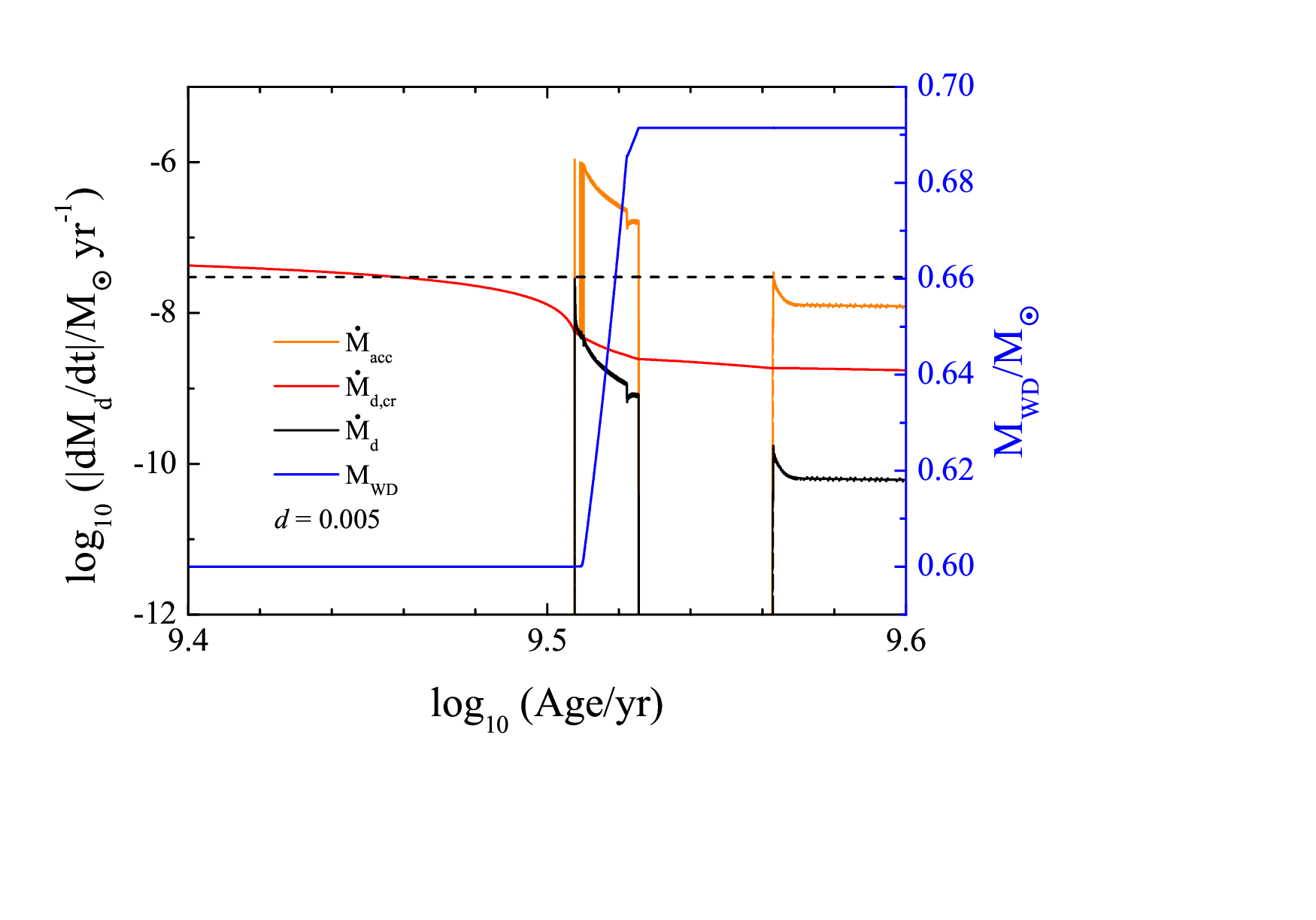}\includegraphics[scale=0.3,trim={2cm 2cm 2cm 0cm}]{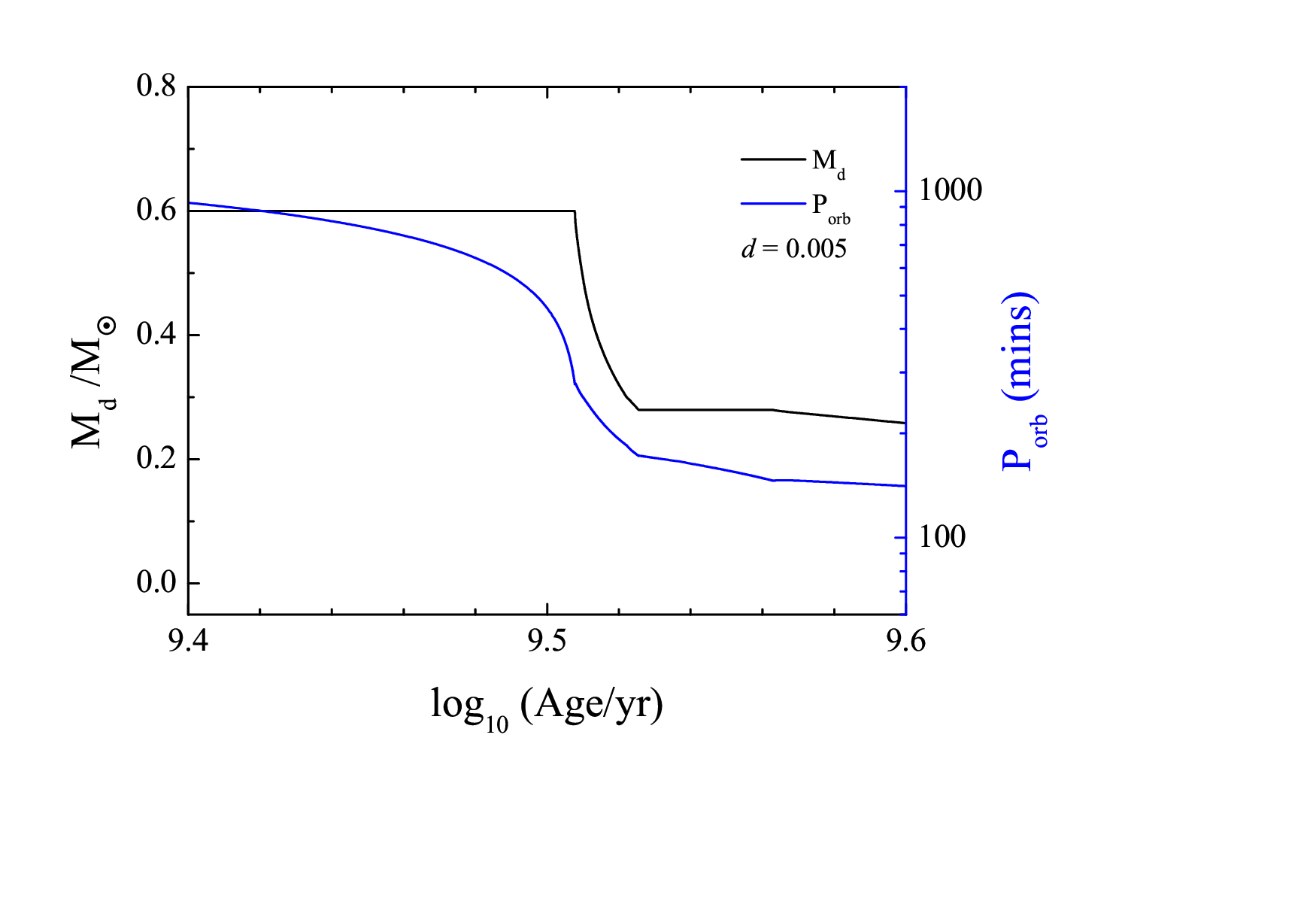}
\includegraphics[scale=0.3,trim={-2cm 0cm 2cm 3cm}]{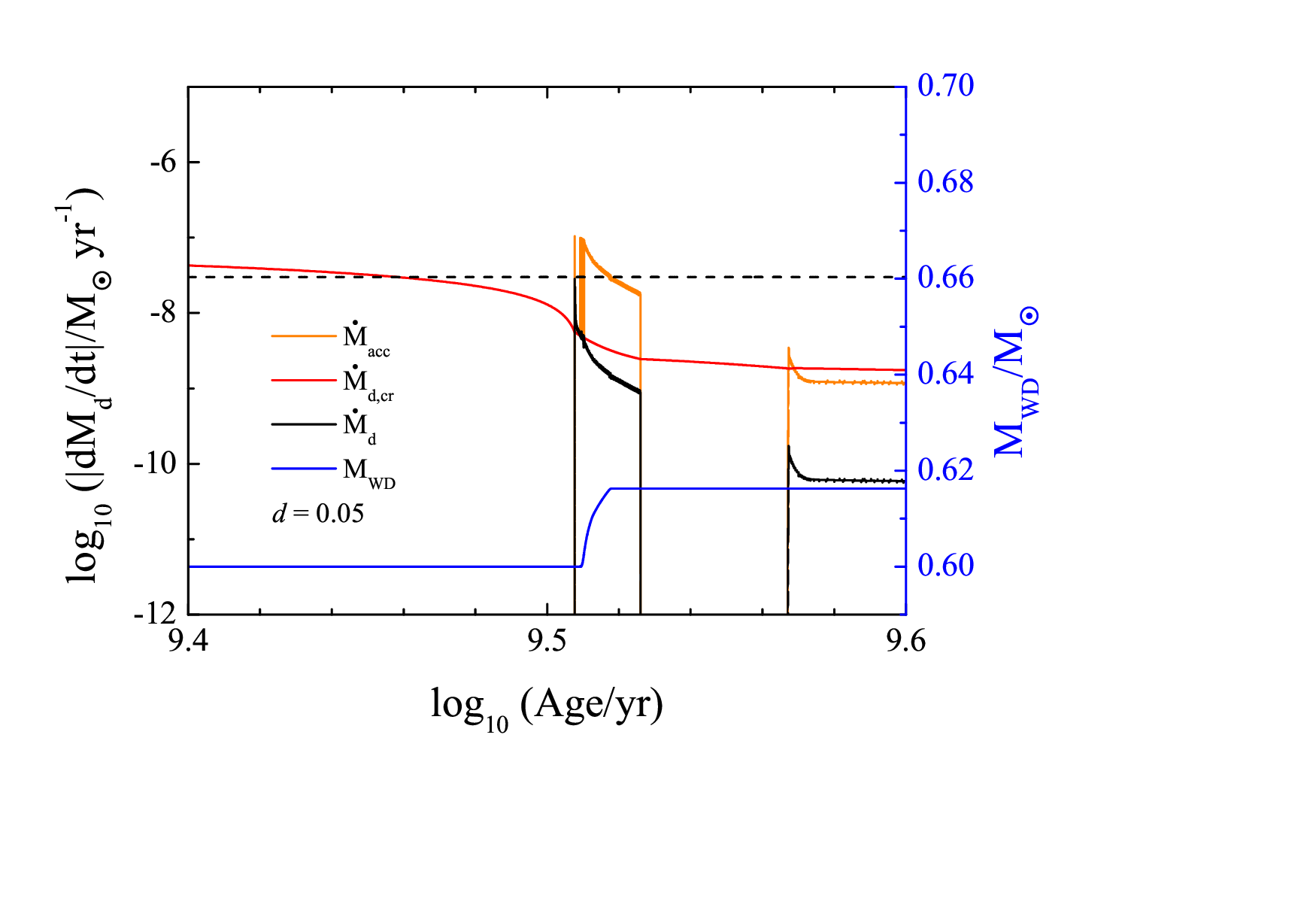}\includegraphics[scale=0.3,trim={4cm 0cm 2cm 3cm}]{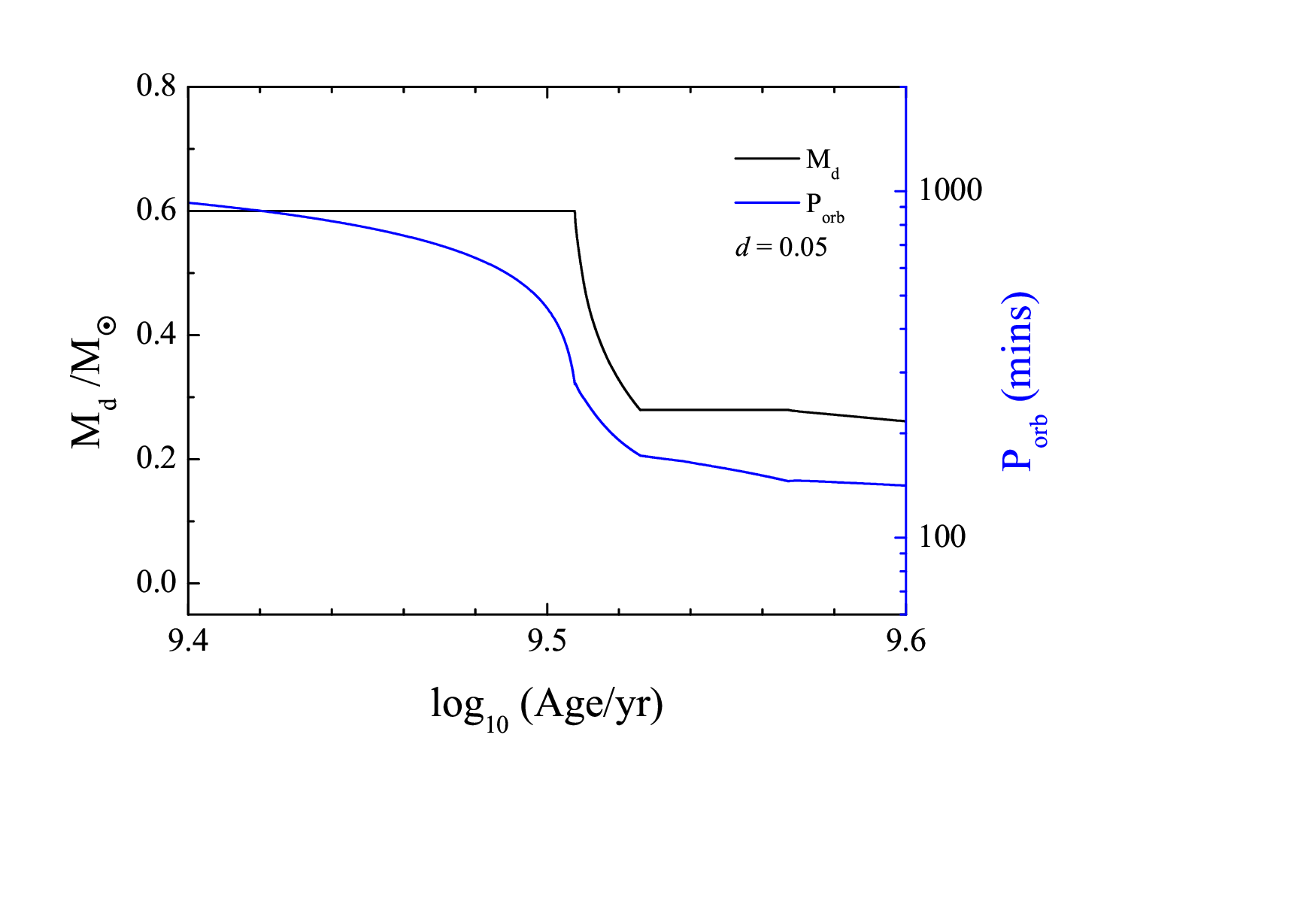}
\vspace{-1cm}
\caption{Same as Figure 1, but with $M_{\rm {d,i}}$ = 0.6 $M_\odot$, $M_{\rm {WD,i}}$ = 0.6 $M_\odot$ and $P_{\rm orb,i}$ = 1.0 d.}
\label{fig:subfig}
\end{figure*}

\begin{figure*}
\centering
\includegraphics[scale=0.30,trim={-2cm 2cm 4cm 0cm}]{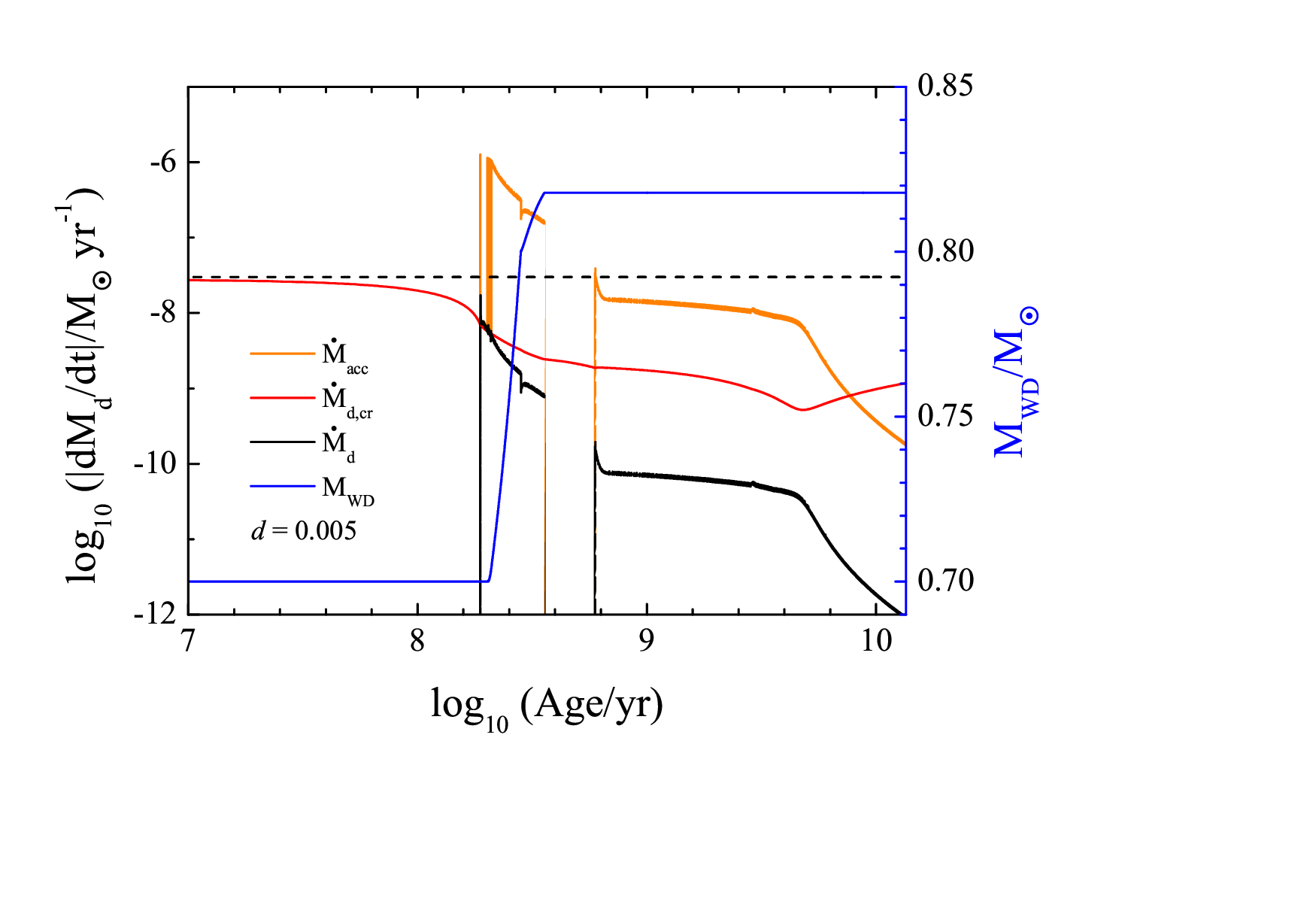}\includegraphics[scale=0.30,trim={2cm 2cm 2cm 0cm}]{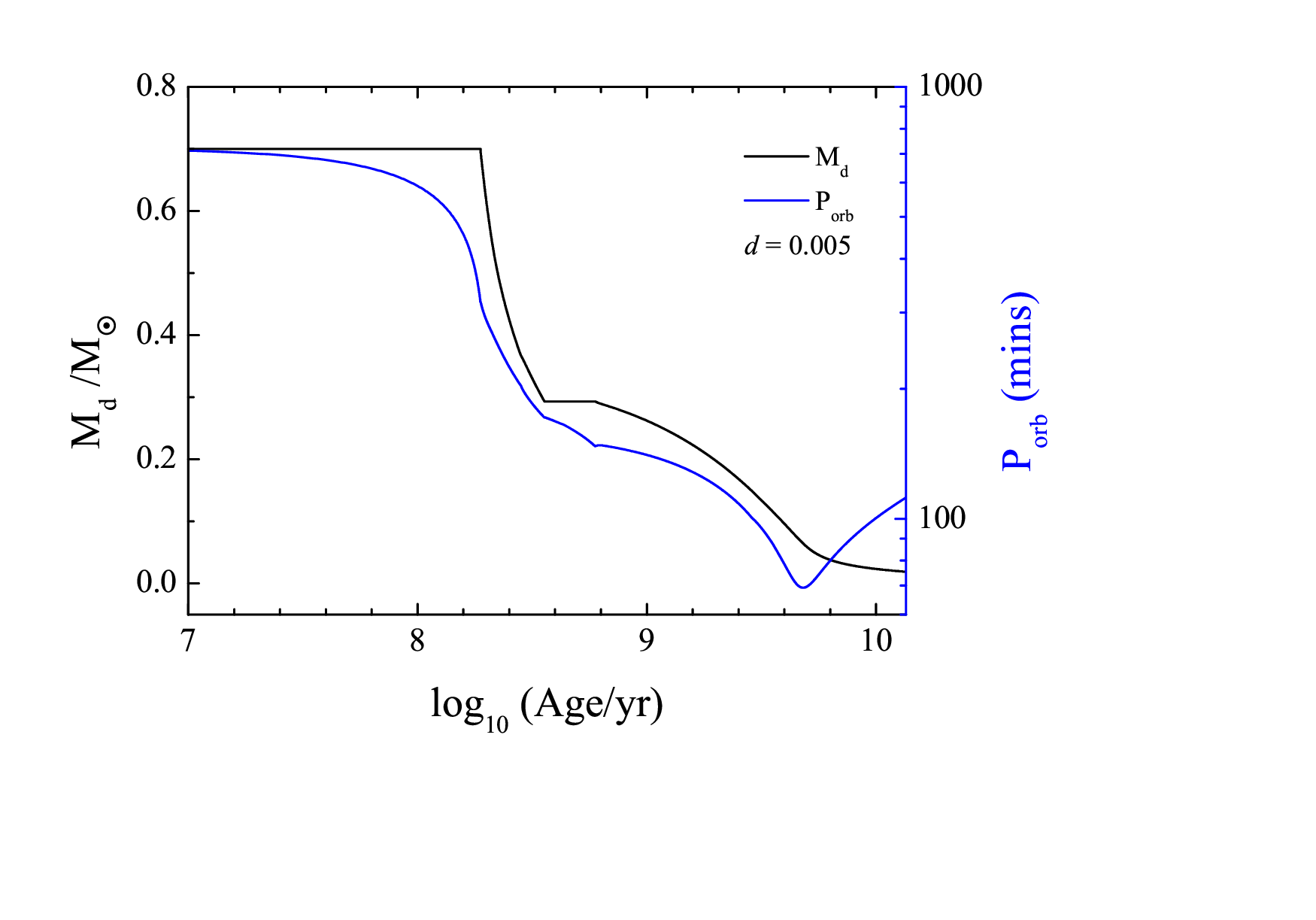}
\includegraphics[scale=0.30,trim={-2cm 0cm 2cm 3cm}]{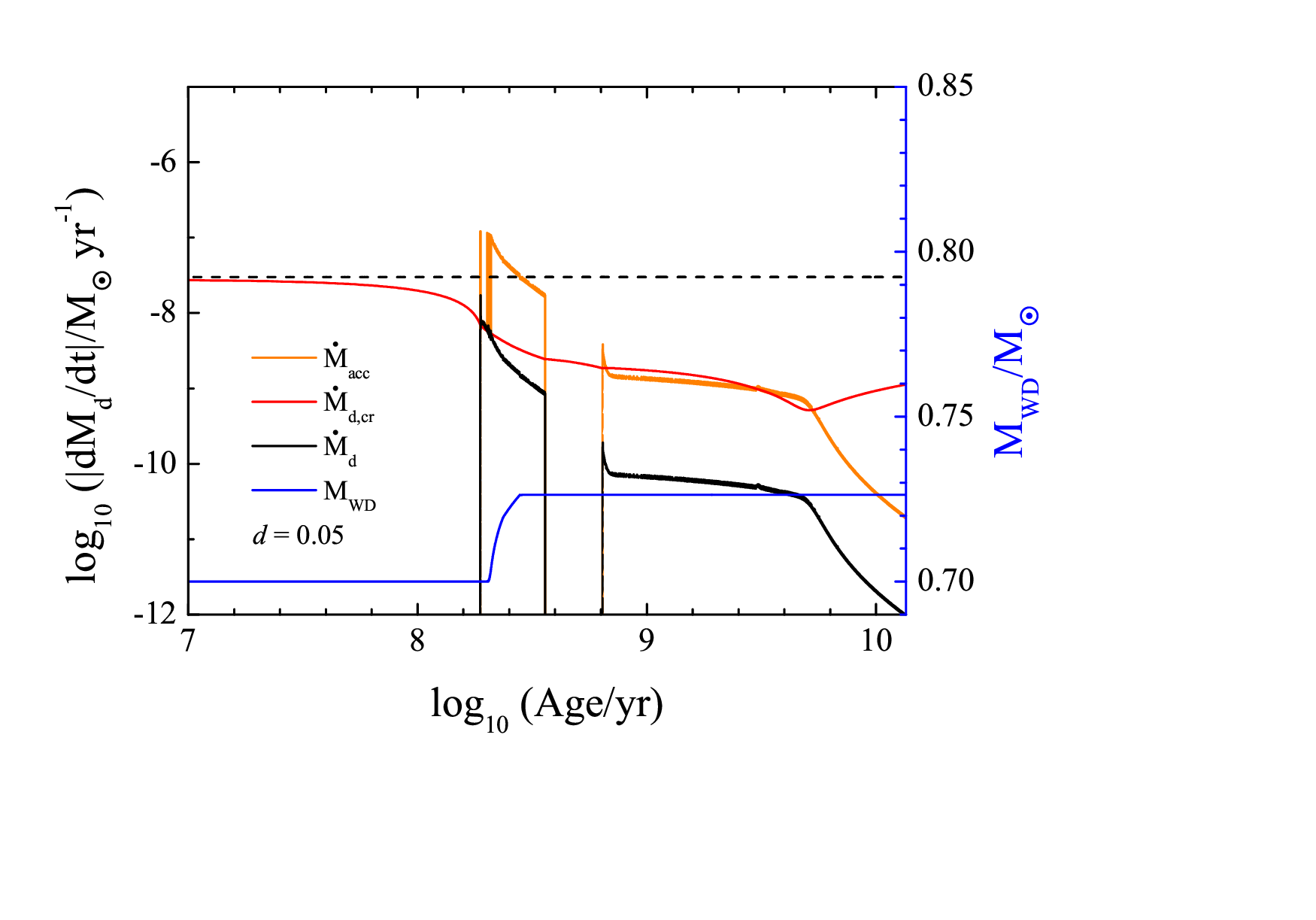}\includegraphics[scale=0.30,trim={4cm 0cm 2cm 3cm}]{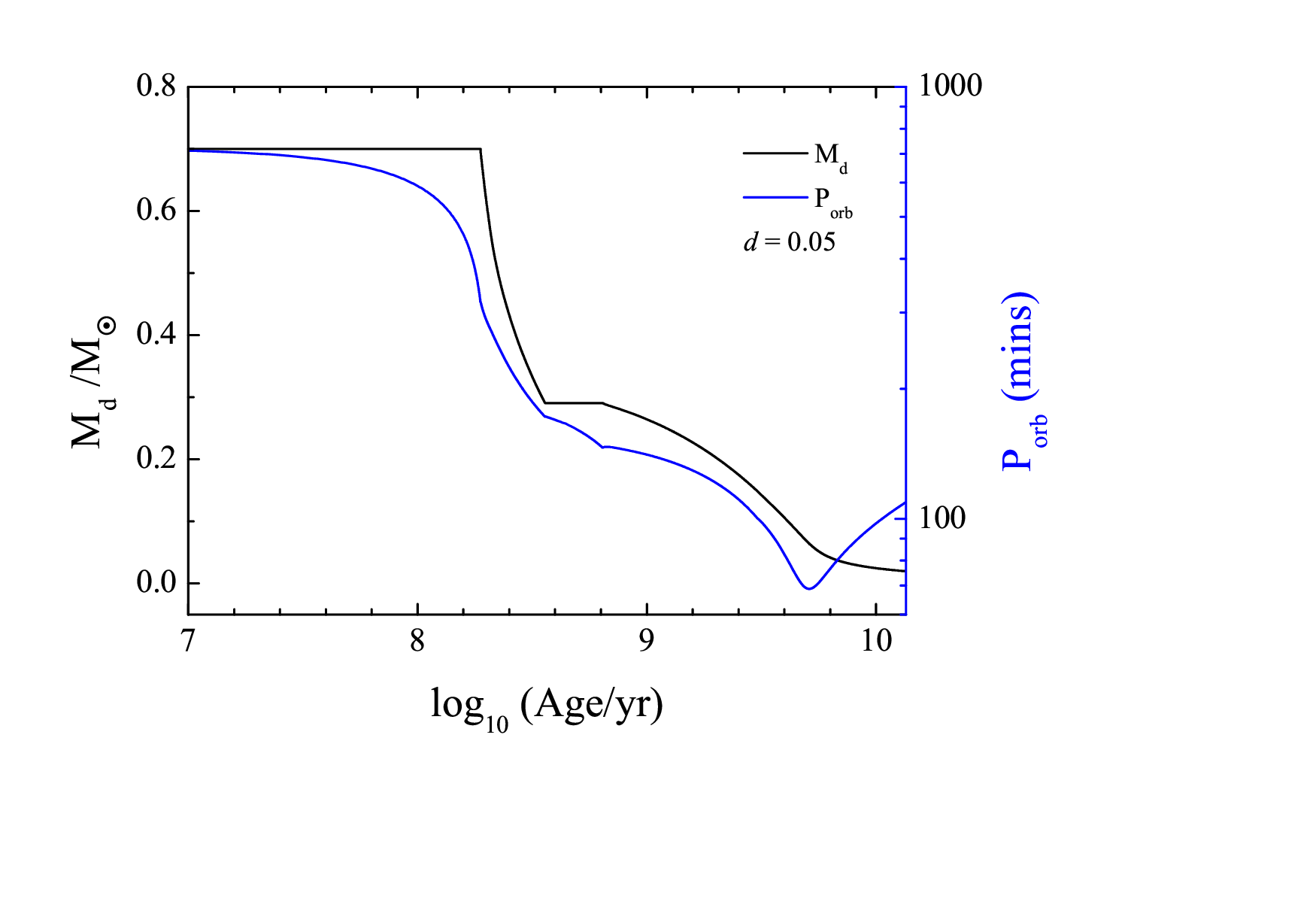}
\vspace{-1cm}
\caption{Same as Figure 1, but with $M_{\rm {d,i}}$ = 0.7 $M_\odot$, $M_{\rm {WD,i}}$ = 0.7 $M_\odot$ and $P_{\rm orb,i}$ = 0.5 d.}
\label{fig:subfig}
\end{figure*}

\begin{figure*}
\centering
\includegraphics[scale=0.30,trim={-2cm 2cm 4cm 0cm}]{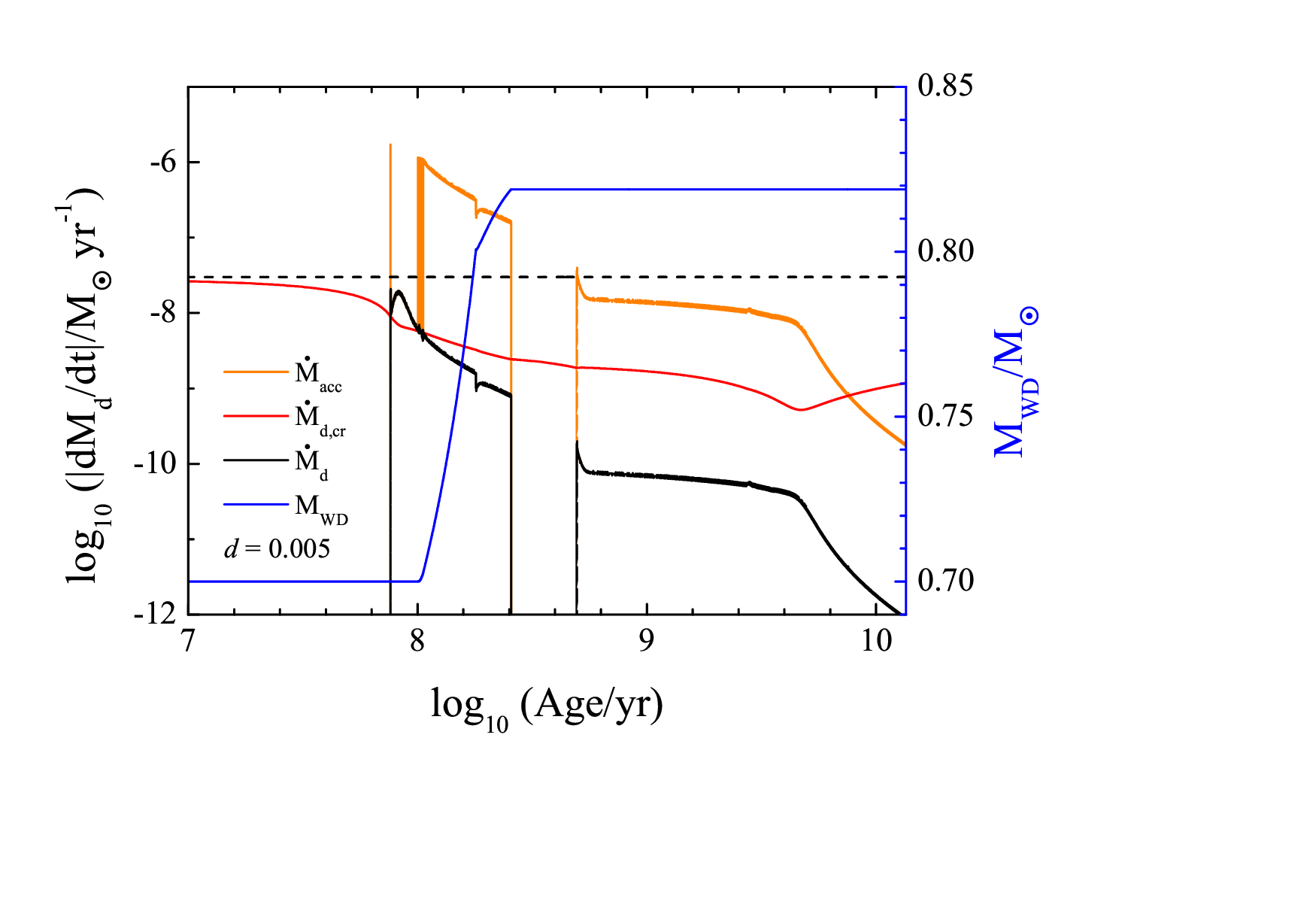}\includegraphics[scale=0.30,trim={2cm 2cm 2cm 0cm}]{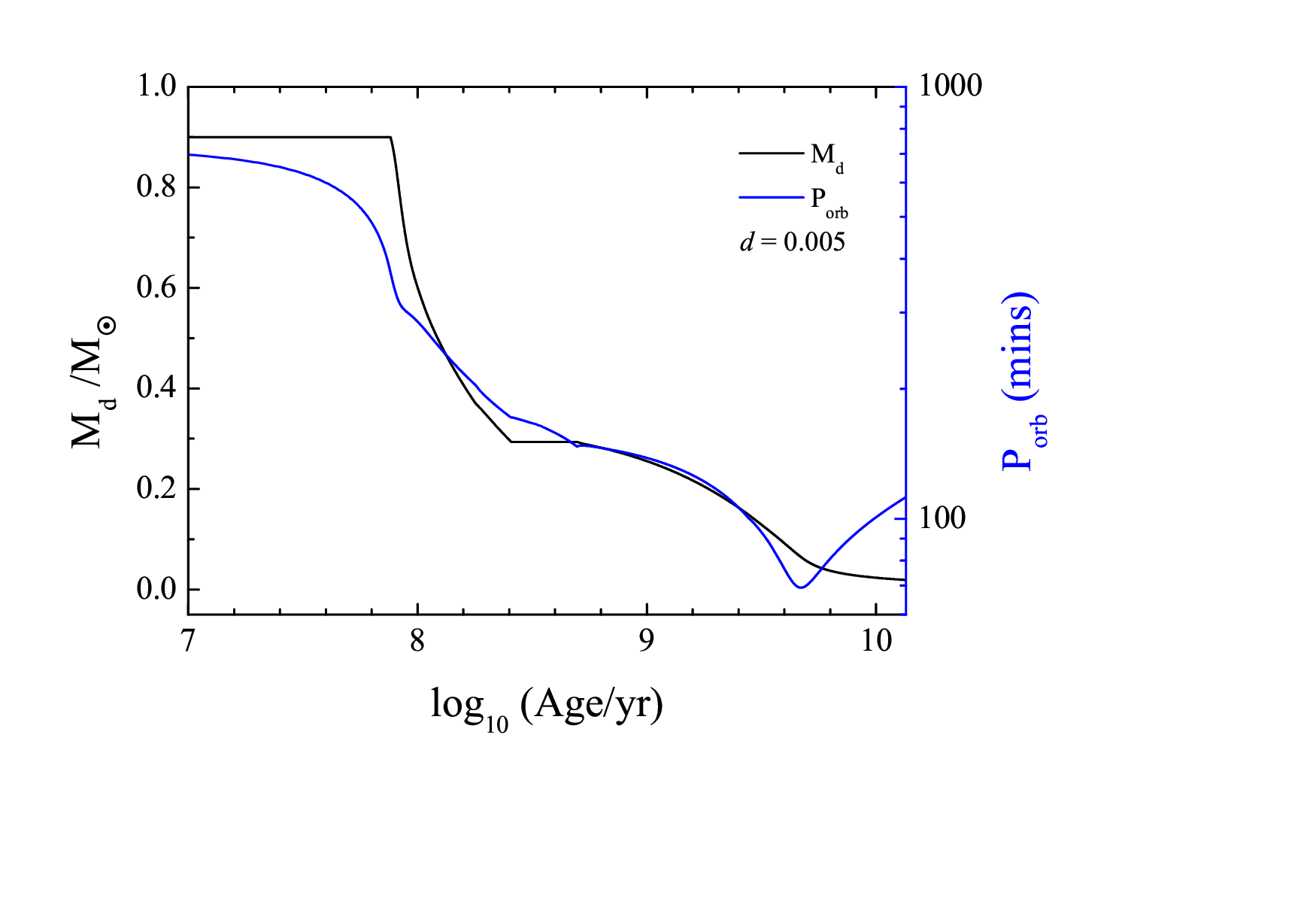}
\includegraphics[scale=0.30,trim={-2cm 0cm 2cm 3cm}]{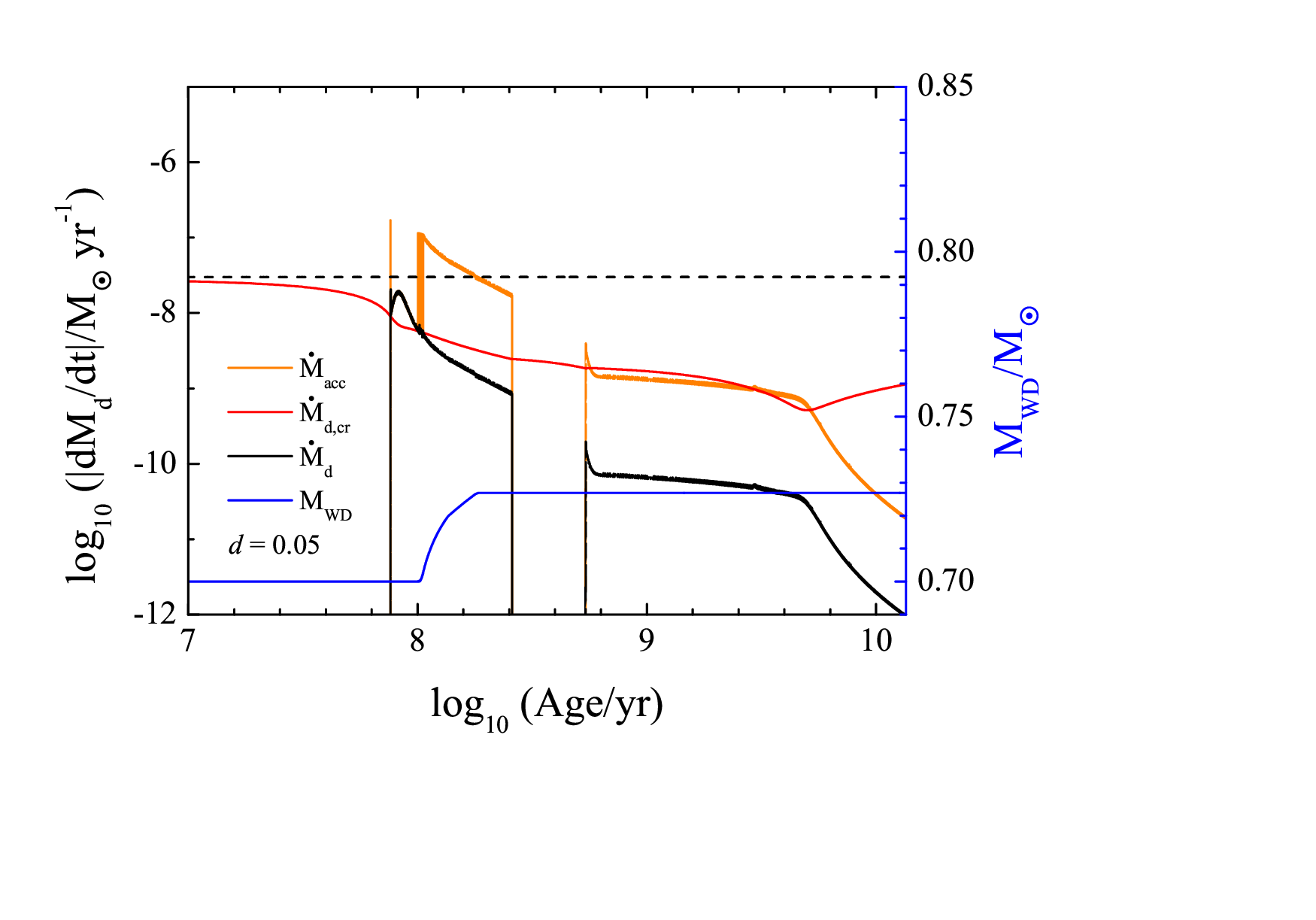}\includegraphics[scale=0.30,trim={4cm 0cm 2cm 3cm}]{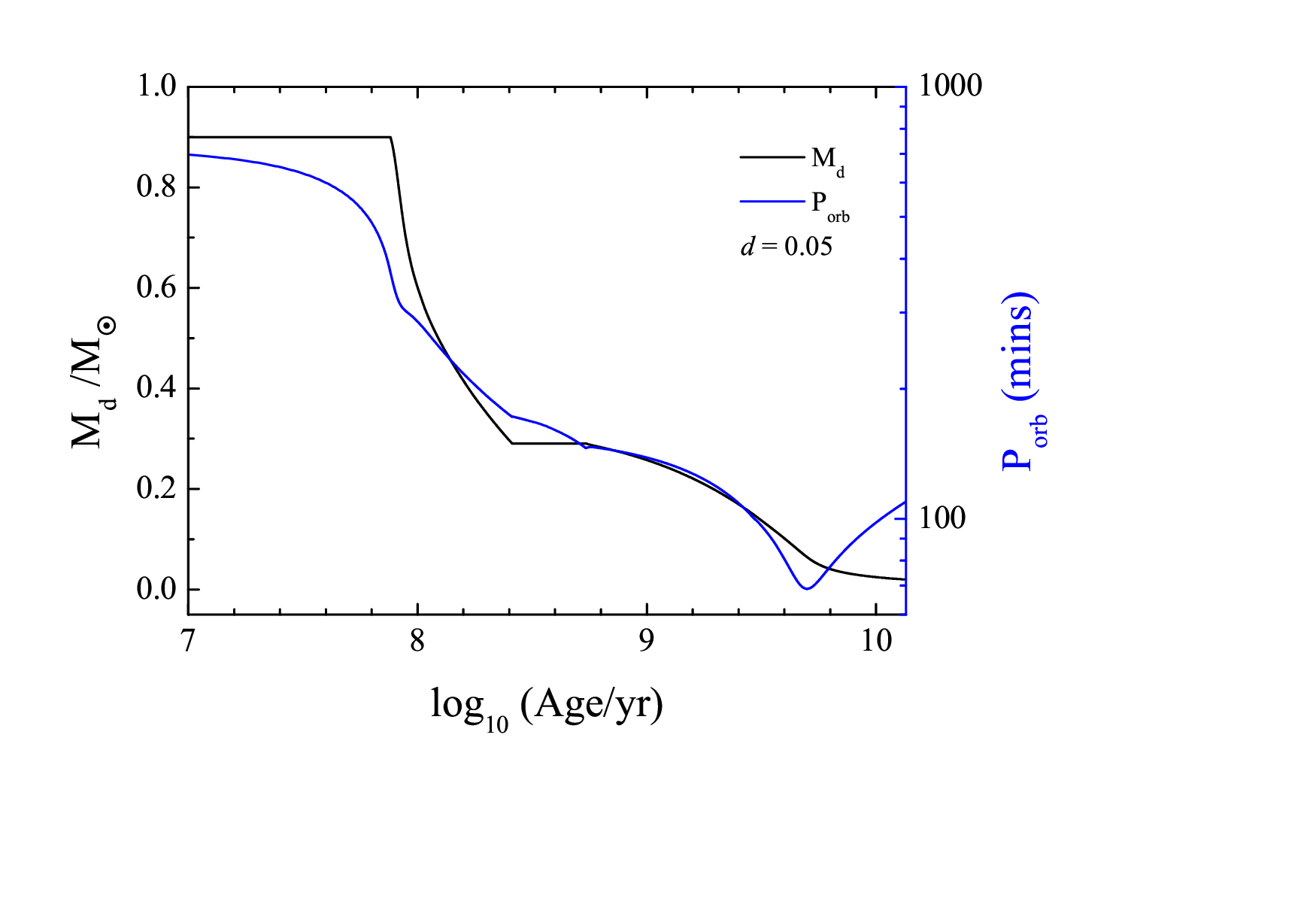}
\vspace{-1cm}
\caption{Same as Figure 1, but with $M_{\rm {d,i}}$ = 0.9 $M_\odot$, $M_{\rm {WD,i}}$ = 0.7 $M_\odot$ and $P_{\rm orb,i}$ = 0.5 d}
\label{fig:subfig}
\end{figure*}

\begin{figure*}
\centering
\includegraphics[scale=0.30,trim={-2cm 2cm 4cm 0cm}]{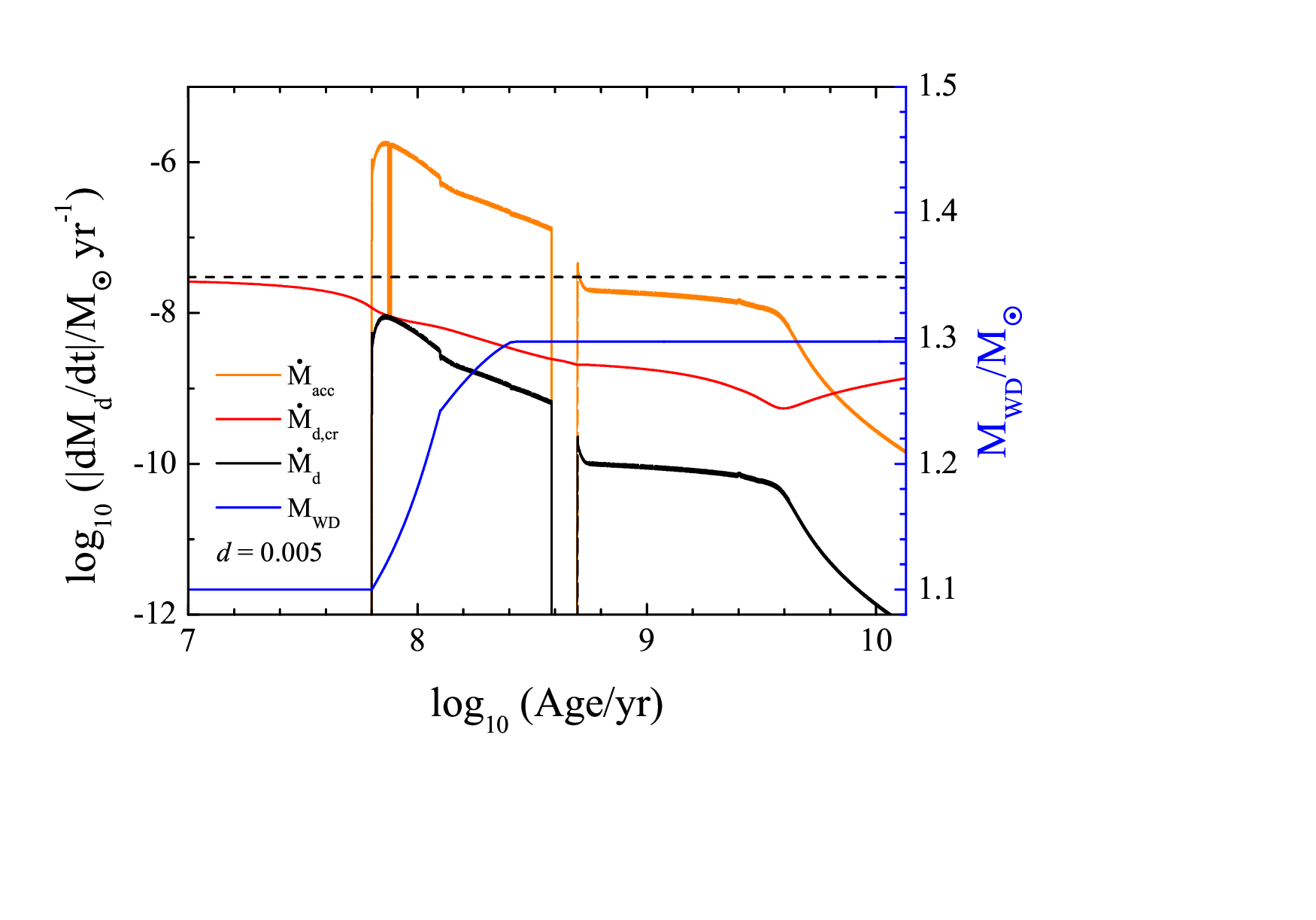}\includegraphics[scale=0.30,trim={2cm 2cm 2cm 0cm}]{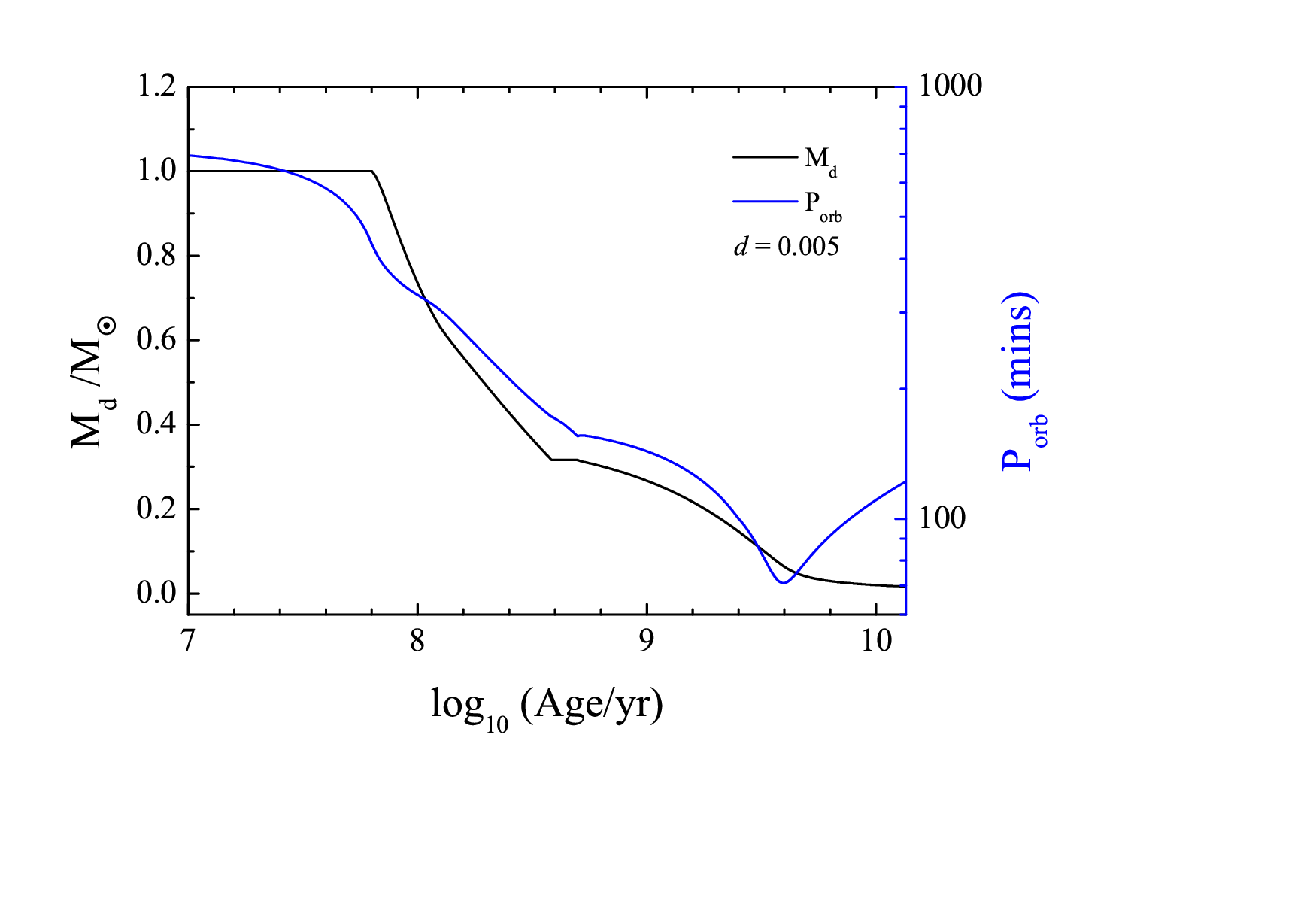}
\includegraphics[scale=0.30,trim={-2cm 0cm 2cm 3cm}]{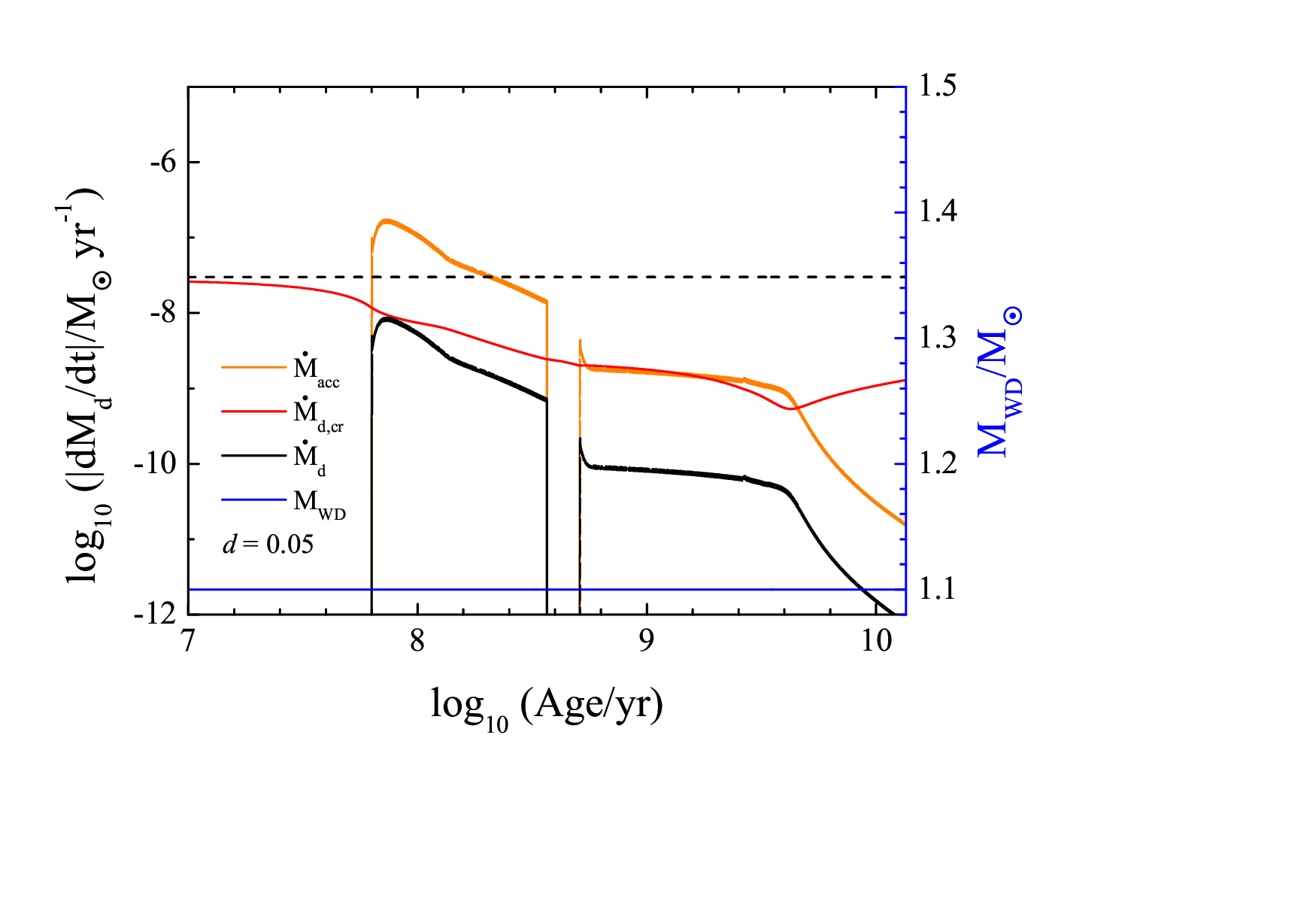}\includegraphics[scale=0.30,trim={4cm 0cm 2cm 3cm}]{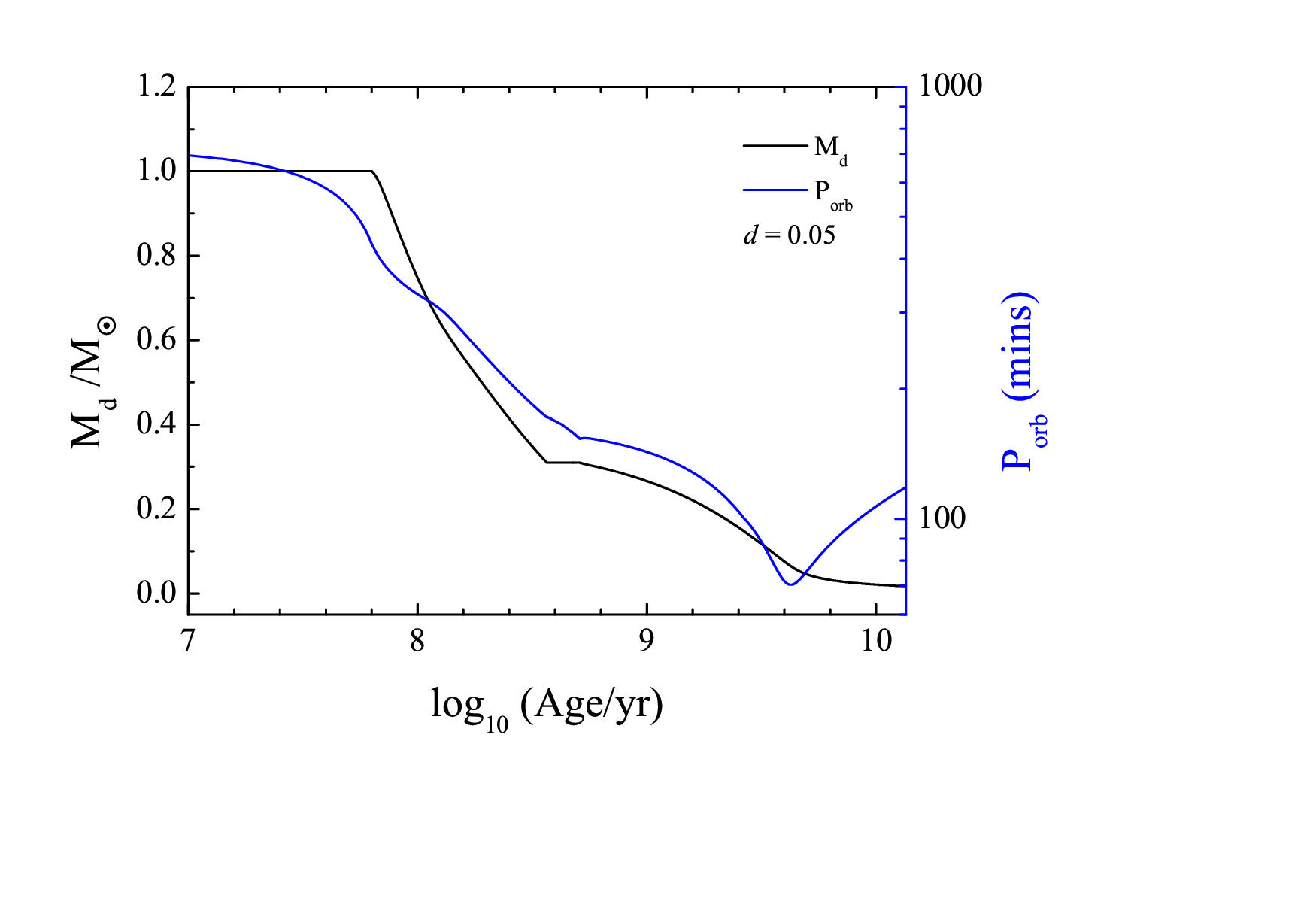}
\vspace{-1cm}
\caption{Same as Figure 1, but with $M_{\rm {d,i}}$ = 1.0 $M_\odot$, $M_{\rm {WD,i}}$ = 1.1 $M_\odot$ and $P_{\rm orb,i}$ = 0.5 d}
\label{fig:subfig}
\end{figure*}

Tables 1, 2, and 3 summarise the initial parameters and the final WD masses for $M_{\rm {WD,i}} =0.6,0.7$, and 1.1 $M_\odot$, respectively. For the same $P_{\rm {orb,i}}$ and a fixed $d$, a high mass ratio $q$ (mass ratio between the donor-star mass to the WD mass) tends to produce a high mass growth of WDs. Similarly, a long initial orbital period $P_{\rm {orb,i}}$ also give rise to a massive WD for a fixed donor-star mass and $d$ because the donor stars with long $P_{\rm {orb,i}}$ experience a deeply nuclear evolution before they fill Roche lobes, resulting in a relatively high mass-transfer rate. Our simulations indicate that duty cycles play vital roles in influencing the mass growth of WDs. A small $d$ tends to produce a massive WD when $d=0.005-0.1$. However, the final WD masses $M_{\rm WD,f}$ with $d=0.002$ are less than those with $d=0.005$. A detailed discussion on this issue see also Section 4. In particular, in Table 3 the WDs with $d = 0.05$, and 0.1 cannot accumulate any matter because these systems experience helium novae stages. Furthermore, almost all WDs cannot increase their masses over $0.01 M_\odot$ if $d=0.1$. Although the actual $d$ may exceed 0.1 observationally \citep{zhang17}. Therefore, the influence of a duty cycle greater than 0.1 on the mass growth of WDs is negligible.

To show more evolutionary details, Figures 1 to 5 present five examples for different initial parameters. Figure 1 plots the evolution of a WD binary with $M_{\rm {d,i}} = 0.6~M_\odot$, $M_{\rm {WD,i}}= 0.6~M_\odot$, and $P_{\rm {orb,i}}=0.5$ d.  At the age of 0.32 Gyr, the donor star fills its Roche lobe, and transfers the surface hydrogen-rich material onto the WD at a rate of
about $10^{-9}-10^{-8}~M_\odot\,\rm yr^{-1}$. Once $\dot{M}_{\rm d} \leq \dot{M}_{\rm d,cr}$, the accretion-disk instability occurs. The evolutionary tracks of the accretion rates of the WD with different $d$ emerge similar profiles (Note that the accretion-rate tracks in all figures only represent evolutionary tendency, and the accreted mass of WDs in a recurrence time is $\Delta M=\dot{M}_{\rm acc} t_{\rm rec}d=\dot{M}_{\rm acc}t_{\rm out}$). This phenomenon originates from that two mass-transfer rates with different duty cycles are approximate same,  while the accretion rates during the outburst state emerge a similar evolutionary law due to $\dot{M}_{\rm acc}=-\dot{M}_{\rm d}/d$.  When $d=0.005$, the accretion rate of the WD is about $2\times10^{-7}-2\times10^{-6}~M_\odot\,\rm yr^{-1}$, which is in the range between $\dot{M}_{\rm H,low}$ and $\dot{M}_{\rm H,up}$, thus the stable hydrogen burning cause the WD mass to increase to be $0.686~M_\odot$. For $d=0.05$, once a relatively low accretion rate of about $2\times10^{-8}-2\times10^{-7}~M_\odot\,\rm yr^{-1}$ decreases to be less than $\dot{M}_{\rm H,low}$, strong hydrogen-shell flash will block the mass growth of the WD. The mass transfer from the donor star divides into two stages: high mass-transfer rate stage, and low mass-transfer rate stage. In the first stage, the mass transfer is driven by the MB. Subsequently, the donor star evolves into a fully convective star (with a mass of about $0.3~M_\odot$), thus the magnetic field lines cannot be locked to the core and magnetic braking cuts off \citep{gros74}. As a result, the mass transfer ceases in the period gap of 2-3 hours \citep{pacz83,spru83}. Under the period gap, the GR drives the mass transfer again at a low rate of about $10^{-10}~M_\odot\,\rm yr^{-1}$. Even if $d=0.005$, the accretion rate is only about $2\times10^{-8}~M_\odot\,\rm yr^{-1}$, which is still less than $\dot{M}_{\rm H,low}$. Therefore, the WD masses keep a constant in the second mass-transfer stage. The right panels shows the evolution of the donor-star mass and the orbital period. It seems that the duty cycles are difficult to influence the evolution of the donor stars and the orbital periods. 

The initial WD mass and the donor-star mass in Figure 2 are same as Figure 1 but for $P_{\rm {orb,i}}$ = 1.0\,d. Because of a long initial orbital period, the mass transfer starts at the age of 3 Gyr. Due to a deeply nuclear evolution of the donor star, a slightly high accretion cause the WD mass to be $0.692~M_\odot$ and $0.616~M_\odot$ for $d=0.005$, and 0.05, respectively. These two WD masses are slightly greater than those ($0.686~M_\odot$ and $0.614~M_\odot$, see also Table 1) with $P_{\rm orb,i}=0.5$ d.

Figure 3 illustrates the evolution of a WD binary with $M_{\rm {d,i}} = 0.7\,M_\odot$, $M_{\rm {WD,i}}= 0.7\,M_\odot$, and $P_{\rm {orb,i}}=0.5$ d. The evolutionary trends are very similar to those of Figure 1 for a same $d$. A high $M_{\rm {WD,i}}$ leads to a high $M_{\rm {H,up}}$, which benefits the mass growth of WDs. Therefore, the mass growth of the WD can reach $0.118~M_\odot$ and $ 0.026~M_\odot$ for $d$ = 0.005, and 0.05, respectively.

To study the influence of the donor-star masses, in Figure 4 we adopt $M_{\rm WD,i}$, and $P_{\rm orb,i}$ same as Figure 3 but for $M_{\rm d,i} = 0.9~M_\odot$. Since the donor-star mass is higher than the WD mass, the mass is transferred to the accretion disk at a rate of $\ga 10^{-8}~M_\odot\,\rm yr^{-1}$ in the early stage of the first mass-transfer stage, which is greater than $\dot{M}_{\rm {d,cr}}$. Without the disc instability, such a mass-transfer rate (is also the accretion rate) is smaller than $\dot{M}_{\rm H,low}$, hence the WD mass cannot increase in the early stage. However, because of a relatively long mass-transfer timescale the WD masses still increase to be 0.819, and $ 0.727~M_\odot$ for $d=0.005$, and 0.05, respectively. The corresponding mass growths are approximately equal to those in Figure 3.

In Figure 5, the initial parameters are $M_{\rm d,i} = 1.0~M_\odot$, $M_{\rm WD,i}= 1.1~M_\odot$, and $P_{\rm orb,i}=0.5~\rm d$. When $d = 0.005$, the disc instability can cause the accretion rate of the WD to remain in the hydrogen-steady-burning region (i.e. $\dot{M}_{\rm H,low}\leq\dot{M}_{\rm acc}\leq \dot{M}_{\rm H,up}$) except for the early mass-transfer stage. Therefore, the accreting material can efficiently accumulate onto the surface of the WD, and the final WD mass is close to $1.30~M_\odot$. On the other hand, when $d = 0.05$, no accreted matter can be retained on the WD surface. The reason is that the low mass accumulating rate for hydrogen burning cannot supply a high accretion rate for stable helium burning, thus helium novae burst. In the early mass-transfer stage, the mass-transfer rate with $d=0.005$ temporarily exceeds $\dot{M}_{\rm d,cr}$ due to AML by the mass loss, resulting in a sudden dip of the accretion rate. However, this phenomenon is absent for $d=0.05$.

\section{Discussion}

In principle, a small duty cycle would produce a high accretion rate during the outburst states. However, a sufficient small duty cycle cannot naturally result in an efficient mass growth of WDs due to the limit of the upper critical mass-accretion rate $\dot{M}_{\rm {H,up}}$. Figure 6 shows the influence of duty cycles on the accretion rate and mass growth of the WD for a binary with $M_{\rm {d,i}}$ = 0.7 $M_\odot$, $M_{\rm {WD,i}}$ = 0.7 $M_\odot$, and $P_{\rm orb,i}$ = 0.5 d. It is clear that $\dot{M}_{\rm {acc}}$ with  $d=0.05$ exceeds the low critical accretion rate (over which the steady H burning is possible) for a short timescale, resulting a tiny mass growth of the WD. Subsequently, strong H-shell flash causes the WD not to accumulate any material because $\dot{M}_{\rm acc}<\dot{M}_{\rm H,low}$. When $d=0.005$, $\dot{M}_{\rm {acc}}$ (about $10^{-6}-10^{-7}{\,M_\odot\,\rm yr}^{-1}$) is always greater than $\dot{M}_{\rm H,low}$ in the first mass-transfer stage, resulting in an efficient mass growth of the WD. For $d=0.002$, $\dot{M}_{\rm acc}$ exceeds $\dot{M}_{\rm {H,up}}$ in the first mass-transfer stage, thus most of the accreted material is lost from the surface of the WD in the form of the optically thick winds. In the second mass-transfer stage, $\dot{M}_{\rm acc}$ is still greater than $\dot{M}_{\rm H,low}$ for a short timescale, while the total mass growth of the WD in two stages is still less than that in $d=0.005$. An efficient mass accumulation on the surface of WDs depends on whether the accretion rates are in the range from $\dot{M}_{\rm H,low}$ to $\dot{M}_{\rm H, up}$, i.e. $\dot{M}_{\rm H,low}\le \dot{M}_{\rm acc}\le\dot{M}_{\rm H, up}$. Therefore, the efficient duty cycle $d$ is related to the mass-transfer rate, and the WD mass. It requires generous numerical calculations to find an efficient $d$, which results in a maximum mass growth of the WD.

Table 4 summarises the efficient duty cycles and the maxima of $M_{\rm {WD,f}}$ for different initial parameters. For a specific WD binary, we adopt different $d$ in steps of 0.001 to find a maximum $M_{\rm{WD,f}}$. Similar to Figure 6, a small $d$ cannot result in the production of a more massive WD.  For initial WD masses of 0.6, 0.7 and 1.1 $M_\odot$, the efficient $d$ are $0.006$ to $0.007$, 0.005, and 0.003. It seems that an initial massive WD corresponds to a small efficient $d$. The reason for this tendency is that the accretion rate of the WD still has chances to be less than the $\dot{M}_{\rm H,up}$ of a massive WD for a small $d$.

\begin{figure*}
\centering
\includegraphics[scale=0.75,trim={0cm 0cm 4cm 0cm}]{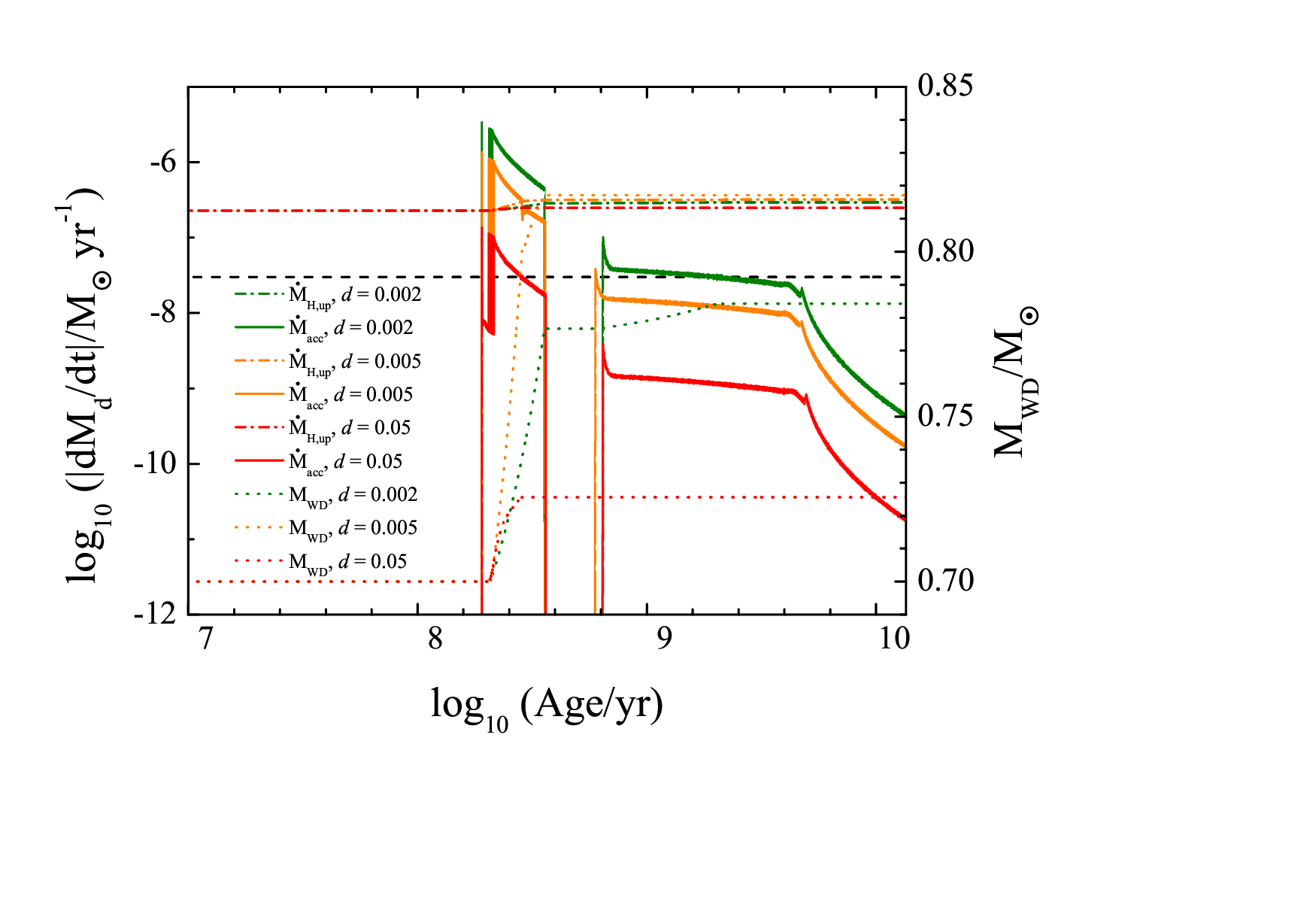}
\vspace{-3cm}
\caption{Same as the left panels of Figure 1, but with $M_{\rm {d,i}}$ = 0.7 $M_\odot$, $M_{\rm {WD,i}}$ = 0.7 $M_\odot$ and $P_{\rm orb,i}$ = 0.5 d. The dashed-dotted curves denote the upper critical accretion rate $\dot{M}_{\rm H,up}$ for steady hydrogen burning. The solid curves, and the dotted curves represent the evolution of the accretion rates of WDs during outburst states, and the WD masses, respectively. The green, orange, and red curves represent the evolutionary tracks when $d=0.05, 0.005$, and 0.002, respectively.}
\label{fig:subfig}
\end{figure*}

In our simulated WD binaries, all WDs cannot increase their masses to the explosion mass of SNe Ia of about $1.38~M_\odot$. However, considering the effect of the disc instability \cite{wang10} obtained the initial parameter space of the progenitors of SNe Ia in the orbital period–donor star mass plane. Especially, several WD binaries with $M_{\rm {d,i}} = 1.6$ to $1.8~M_\odot$, $M_{\rm {WD,i}}= 0.61\,M_\odot$, and $P_{\rm {orb,i}}=1.2$ to $1.6$ d are potential progenitor of SNe Ia. The discrepancy between our results and those by \cite{wang10} should arise from different initial parameters including donor-star masses and orbital periods. To achieve an efficient mass growth of the surface of WDs, the initial donor-star masses of the progenitors of SNe Ia must have masses greater than $1.1~M_\odot$ \cite[see also][]{wang10}, which ensures a thermal-timescale mass transfer for a long time. However, our work aims at the mass growth of WDs in DNe, thus the maximum donor-star mass is $1.0~M_\odot$. Furthermore, \cite{wang10} took a constant duty cycle as $d=0.01$ in their simulations. However, in this work we study the influence of different duty cycles on the mass growth of WDs.

Our calculations indicate that the growth of a WD mass can reach a maximum of around 0.1 to 0.15 $M_\odot$ for $M_{\rm WD,i}=0.6$ to $0.7 ~M_\odot$ during the disc instability. The extent of this increase, for low $M_{\rm {WD,i}}$ might alleviate the mass discrepancy between single WDs and WDs in CVs \citep{zor11,liu16}. However duty cycles of DNe remains uncertain while the birthrate of DNe amongst all CVs is also poorly known. So it is hard to draw a robust conclusion as to whether DNe can solve this WD mass problem.

We also find that the evolution of orbital periods and donor-star mass is hardly affected by duty cycles. This is consistent with a known relation between the donor-star mass and the orbital period. The main reason is that the timescale of the outburst states is much shorter than the recurrence time, and the secular AML due to mass loss is smaller than the contribution of MB. Therefore, DIMs and different duty cycles can mainly influence the WD masses.

\section{Conclusion}
In this paper, we explore the influence of the disc instability on the WD-mass growth in DNe. Employing the stellar evolution code MESA, we model the evolution of dozens WD binaries consisting of a WD and a MS companion for different duty cycles. Our main results are summarised as follows.

(1) The DIM plays an important role in influencing the mass growth of WDs during DNe. For $M_{\rm {WD,i}}$= 0.6, 0.7 and 1.1 $M_\odot$, the maximum mass growth are 0.1, 0.13 and 0.21 $M_\odot$, respectively. Therefore, the DIM in DNe could alleviate the WD mass problem to some extent.

(2) Duty cycles are key factors in determining the mass growth of WDs. For a specific WD binary, there exists an efficient $d$, at which the mass growth of WDs reaches a maximum. This efficient $d$ is related to the initial parameters including $M_{\rm {WD,i}}$, $M_{\rm {d,i}}$, and $P_{\rm {orb,i}}$. In our simulated parameter space, for $M_{\rm {WD,i}} = 0.6, 0.7$, and 1.1 $M_\odot$ the efficient duty cycles locate at 0.006$\,\leq$$d$$\,\leq$0.007, $d$\,=\,0.005 and $d$\,=\,0.003, respectively.

(3) For the same $d$, a long initial orbital period or a high donor-star mass leads to a large mass growth of WDs. This is why \cite{king03} favoured the long-period DNe as an alternative channel of SNe Ia. Within our simulated parameter space, no WDs can accumulate their masses to the explosion mass of SNe Ia of 1.38 $M_\odot$, thus no SNe Ia happen.

(4) The final donor-star masses and orbital periods are difficult to be affected by $d$. The main reason is that the timescale of the outburst state is much shorter than the recurrence time, resulting in the AML due to mass loss during the hydrogen and helium unstable burning is much smaller than the contribution by MB.

(5) A canonical duty cycle of $d$ = 0.01 used in the previous studies for the progenitors of SNe Ia \citep[see also][]{wang10}. may overestimate or underestimate the effect of DIMs, and the latter seems more likely. It requires a relatively small duty cycle of $d=0.003-0.007$ in order to alleviate the WD mass problem.

In the future, we plan to explore observational studies of some DNe \citep[e.g. EY Cyg1954+3221 and RU Peg2214+1242,][]{rit03}. These DNe possess relatively high WD masses, however, their outburst details are not well understood. We expect to constrain the duty cycles and some key parameters for these sources, and reproduce their evolutionary history.

\section*{Acknowledgements}

We are grateful to the referee, Professor Christopher Tout for the valuable comments which helped improve this manuscript. This work was supported by the Natural Science Foundation of China under grant Nos. U2031116, 12273014, 12041301, 12121003, and U1838201.

\section*{Data Availability}
All data underlying this article will be shared on reasonable request to the corresponding author.










\bsp	
\label{lastpage}
\end{document}